\documentclass[prb,aps,preprint]{revtex4}
\usepackage{graphicx}
\usepackage{dcolumn}
\usepackage{bm}
\begin{document}
\title {Microscopic origin of charge impurity scattering and flicker noise in MoS$_2$ field effect transistors}
\author{Subhamoy Ghatak$^1$, Sumanta Mukherjee$^2$, Manish Jain$^1$, D. D. Sarma$^2$ and Arindam Ghosh$^1$}
\address{$^1$Department of Physics, Indian Institute of Science, Bangalore 560 012, India}
\address{$^2$Solid State Structural Chemistry Unit, Indian Institute of Science, Bangalore 560 012, India}

\begin{abstract}

Scattering of charge carriers and flicker noise in electrical transport are the central performance limiting factors in electronic devices, but
their microscopic origin in molybdenum disulphide~(MoS$_2$)-based field effect transistors remains poorly understood. Here, we show that both
carrier scattering and low-frequency $1/f$ noise in mechanically exfoliated ultra-thin MoS$_2$ layers are determined by the localized trap
states located within the MoS$_2$ channel itself. The trap states not only act as Coulomb scattering centers that determine transport in both
equilibrium ($eV< k_BT$) and non-equilibrium ($eV>k_BT$) regimes, where $V$ and $T$ are the source drain bias and temperature respectively, but
also exchange carriers with the channel to produce the conductivity noise. The internal origin of the trap states was further confirmed by
studying noise in MoS$_2$ films deposited on crystalline boron nitride substrates. Possible origin and nature of the trap states is also
discussed.
\end{abstract}


\maketitle

Atomically thin films of MoS$_2$ have emerged as a promising platform for transparent flexible electronics. In the field effect geometry,
MoS$_2$ offers several advantages that include large on-off ratio, immunity against short channel effects, and small subthreshold
swing~\cite{single,channellengthscaling}. These promise MoS$_2$-based logic devices~\cite{kisintegratedcircuit,integratedbilayer} and
energy-efficient field effect transistor~\cite{Yin,KallolNatNano,kallolSSC}, but in the generic back gated geometry, the electron mobility of
MoS$_2$ field effect devices (MoS$_2$~FET) is generally poor ($\lesssim 20$~cm$^2$/Vs), which may restrict its application as fast transistors
and rf devices~\cite{howgood}. Since the phonon-limited room temperature mobility~\cite{phononlimitedmobility} in MoS$_2$ can be as large as
$\sim 400$~cm$^2$/V.s, the factors that restrict carrier mobility in MoS$_2$ are of great research interest~\cite{scandium}. Recently, some
studies have reported to achieve higher mobility in top gated devices with a high-$k$ dielectric~\cite{single,dualgatemos2ieee,cnrraomos2}. This
suggest that charged defects play a crucial role in determining the performance of MoS$_2$-based transistors, but the microscopic origin of
these defects has remained unclear.

Disorder in ultra-thin MoS$_2$ transistor may arise due to both external and internal factors. The former, which primarily involves
substrate-bound charge traps, roughness or adsorbates, was shown to affect both electron mobility and flicker noise in graphene
FETs~\cite{corrugation,Adamchargedimpuritysolidstatecomm,chargedimpuritygraphene,ouracsnano}. Although a similar picture has been suggested for
MoS$_2$ as well~\cite{cnrraomos2,nanopatch,natureofelectronic}, the occurrence of strong localization even near room temperature and at high
carrier density indicate a much stronger disorder in MoS$_2$. Among the internal factors, sulphur vacancies are often discussed which can
introduce localized donor states inside the bandgap and determine the electrical transport properties~\cite{quisulphurvac}. It has been reported
that presence of atmospheric oxygen and water vapour can lead to oxidation of MoS$_2$ resulting in MoO$_3$~\cite{MoO3raman}. Moreover,
nucleation of metallic 1T polytrope and the grain boundaries can also act as structural disorder as observed in chemically exfoliated
MoS$_2$~\cite{photolumichemicallyexfo}.

To address the origin of performance-limiting disorder, we have carried out electrical transport measurements combined with
time dependent conductance fluctuations {\it i.e.} $1/f$ noise. The purpose of employing 1/$f$ noise stems from its sensitivity to both
structural inhomogeneity in solids,~\cite{Chandnidiprl} and charge traps in proximity to the channel as shown for
graphene~\cite{ouracsnano,atindaapl,atindaprl} and silicon~FETs~\cite{jayaraman,loannidis}. In our experiments, carrier density dependence of
both conductivity and noise measurements suggest that disorder is dominated by trapped charges. However, the trap density calculated from both
turned out two orders of magnitude higher than the SiO$_2$ surface trap density, suggesting that the traps are internal to bulk of MoS$_2$
films. This observation was also supported by 1/$f$ noise measurement in MoS$_2$ devices on trap-free hexagonal boron nitride~(hBN) substarte.

\begin{figure*}
\begin{center}
\includegraphics[width=0.7\linewidth]{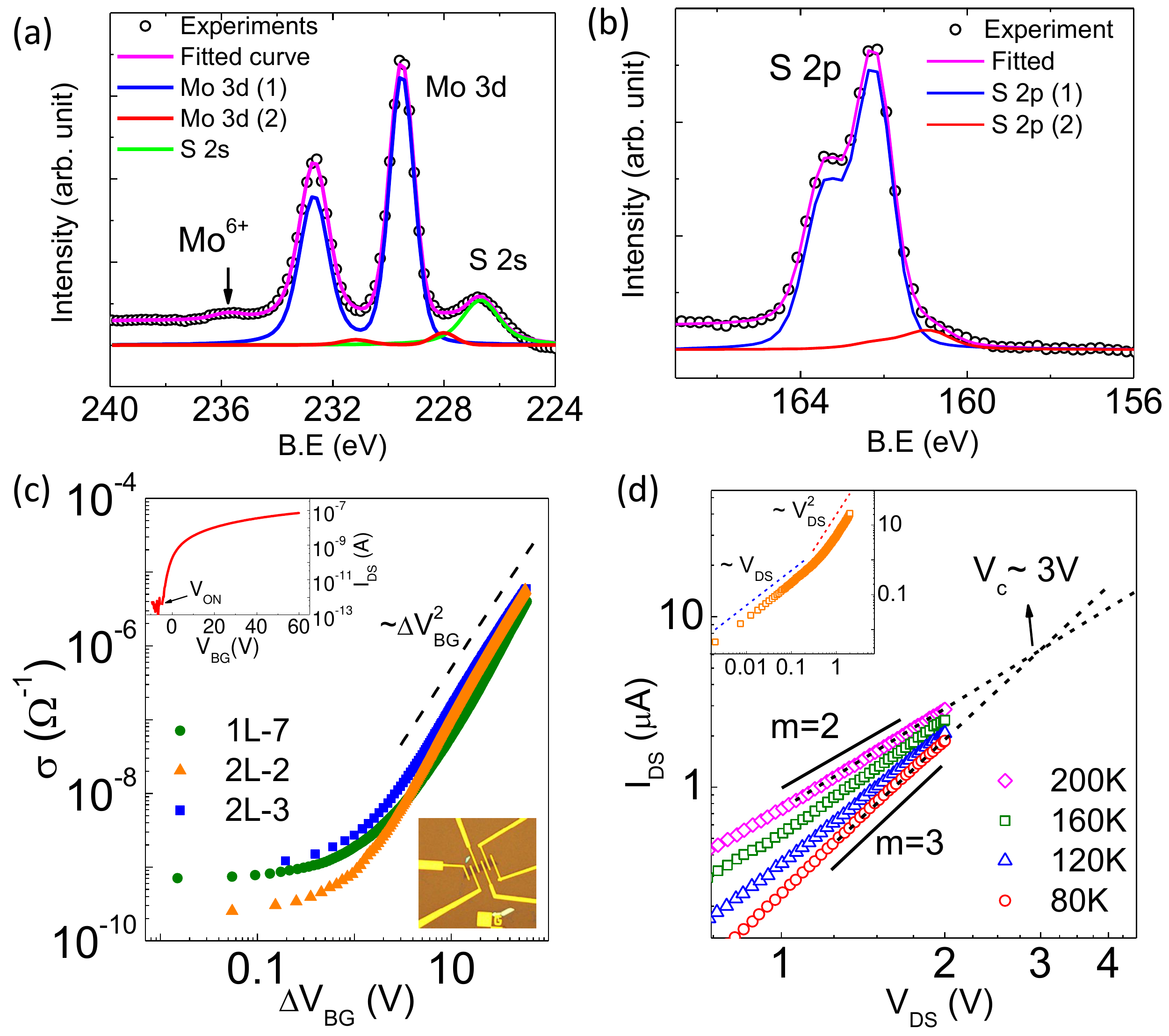}
\caption{(a)~Deconvolution of the experimental photo electron spectra~(black open circles) of Mo 3d to obtain different crystallographic
components namely 2H phase (blue line). The red line indicates presence of additional phase (see text). The solid orange line is the overall
simulated spectra. (b)~Similar deconvolution of S 2p spectra. (c)~Conductivity showing a $\Delta V_{BG}^2$ dependence at high $T$ and $V_{BG}$
for three different devices. Inset: Transfer characteristics of a single layer MoS$_2$ device on SiO$_2$ substrate with
$V_{DS}~=~10~$mV~(top-left), and Optical micrograph of a single layer MoS$_2$ device on Si/SiO$_2$ wafer with 300~nm oxide. The scale bar is
10~$\mu$m~(bottom-right). (d)~$I_{DS}-V_{DS}$ measurement of the 1L-8 device at high $V_{DS}$ and $T~<~200$~K. Inset: Ohmic and non-Ohmic
regions of the $I_{DS}-V_{DS}$ characteristic at 300~K. }
\end{center}
\end{figure*}

Bulk MoS$_2$ crystals were obtained from SPI supplies and were at first thoroughly characterized with X-ray photoemission spectroscopy~(XPS) and
Raman spectroscopy~(See Fig.~S1e in supplementary material). Mo 3d and S 2p XPS spectra obtained from the bulk crystal are shown as open circles
in Fig.~1a and 1b along with various component spectra obtained from spectral decomposition. The major contribution in both Mo 3$d$ and S 2$p$
peaks arises due to 2H phase~(blue lines) of MoS$_2$~\cite{photolumichemicallyexfo}, although additional features~(red lines) at lower binding
energies were observed in both spectra at 4\% intensity ratio~(see supplementary material for details). This suggests presence of structural
inhomogeneity in bulk MoS$_2$ crystals. For electrical characterization, single and bilayer MoS$_2$ flakes were exfoliated from bulk crystals on
300 nm Si/SiO$_2$ wafer using the scotch tape technique. Contact pads were designed using standard ebeam lithography and metallization~(see
experimental details in supplementary material). An optical micrograph of a typical device is shown in bottom-right inset of Fig.~1c.
Two-probe conductance measurement was adapted due to high resistance of the samples. In all devices~(see TABLE~I in supplementary material), the
current-voltage~($I_{DS}-V_{DS}$) characteristics were linear around room temperature at high back gate voltage~($V_{BG}$) and low source-drain
bias~$V_{DS} \leq$~100~mV (see Fig.~S5 in supplementary material). The characteristics became slowly non-linear as temperature was decreased.
Both conductivity($\sigma$) and $1/f$-noise in current were measured in the linear current-voltage regime. This restricted the excitation bias
to $V_{DS}\leq$~10~mV. Although in 2-probe measurement contact resistance can not be avoided, our devices shows much higher total resistance
than typical contact resistance~($R_c$), reported recently by Transmission Line Measurement~(TLM) study for Au contacted MoS$_2$ devices with
similar transfer characteristics~\cite{channellengthscaling}.


The backgate transfer characteristics of a typical single layer device at 300~K is shown in top-left inset of Fig.~1c. The conduction occurred
at positive $V_{BG}$ indicating an intrinsic $n$-doping of the channel. We defined $V_{ON}$~(top-left inset of Fig.~1c) as the backgate voltage
at which the current through the device became measurable~($>$~10$^{-12}$~A) at $T$~=~300~K. The parameter $\Delta V_{BG} = V_{BG} - V_{ON}$, is
then approximately proportional to carrier density~($n$) particularly at high $\Delta V_{BG}$. In Fig.~1c, we have plotted conductivity as
function of $\Delta V_{BG}$ for three different devices and found $\sigma \propto \Delta V_{BG}^2$ for $\Delta V_{BG} \geq 6$~V. This indicates
that the charge transport is dominated by unscreened Coulomb impurity scattering~\cite{sigmansqdassarma}. Such $\Delta V_{BG}^2$ dependence was
observed in \emph{all} the devices around 300~K.

To explore the microscopic origin of the charged impurities or traps, we perform $I_{DS}-V_{DS}$ measurements at high source-drain bias, which
is a sensitive tool to investigate the disorder configuration in organic thin film transistors~\cite{Berleb,mottgurneymorpugo} and reduced
graphene oxide~\cite{khandekerrgo}. At low source-drain bias $I_{DS} \propto V_{DS}$, although it deviated from linearity as $V_{DS}$ was
increased leading to $I_{DS} \propto V_{DS}^m$ with m$~\geq~$2 particularly below 200~K~(inset of Fig.~1d). The $I_{DS}-V_{DS}$ characteristics
in our devices were highly symmetric in the entire drain-source voltage range which eliminates dominance of Schottky barrier at the contact~(see
Fig.~S5 in supplementary material). Therefore, we attribute this to trap-assisted space charge limited conduction~(SCLC) in MoS$_2$ devices and
hence presence of an exponential distribution of trap states~\cite{mottgurneyourAPL}. An exponential distribution can not originate from
SiO$_2$, indicating a bulk or structural origin of these trap states which also leads to charged impurity scattering when occupied. In SCLC
regime, the critical voltage~($V_c$) where $I_{DS}-V_{DS}$ curves at different temperature intersect each other~\cite{mottgurneykumer}~(Fig.~1d)
provides an estimation of volume trap density $N_T$. $N_T$ is related to $V_c$ by the relation $N_T=~2\varepsilon_r \varepsilon_0 V_c/qL^2$. We
obtained $V_c \approx 3$V by extrapolating the $I_{DS}-V_{DS}$ characteristics and found $N_T \sim 2\times 10^{17}$~cm$^{-3}$ for 1L-8 device,
where $\varepsilon_r~\approx~3.3$ is the relative permittivity of MoS$_2$.

\begin{figure*}
\begin{center}
\includegraphics[width=0.7\linewidth]{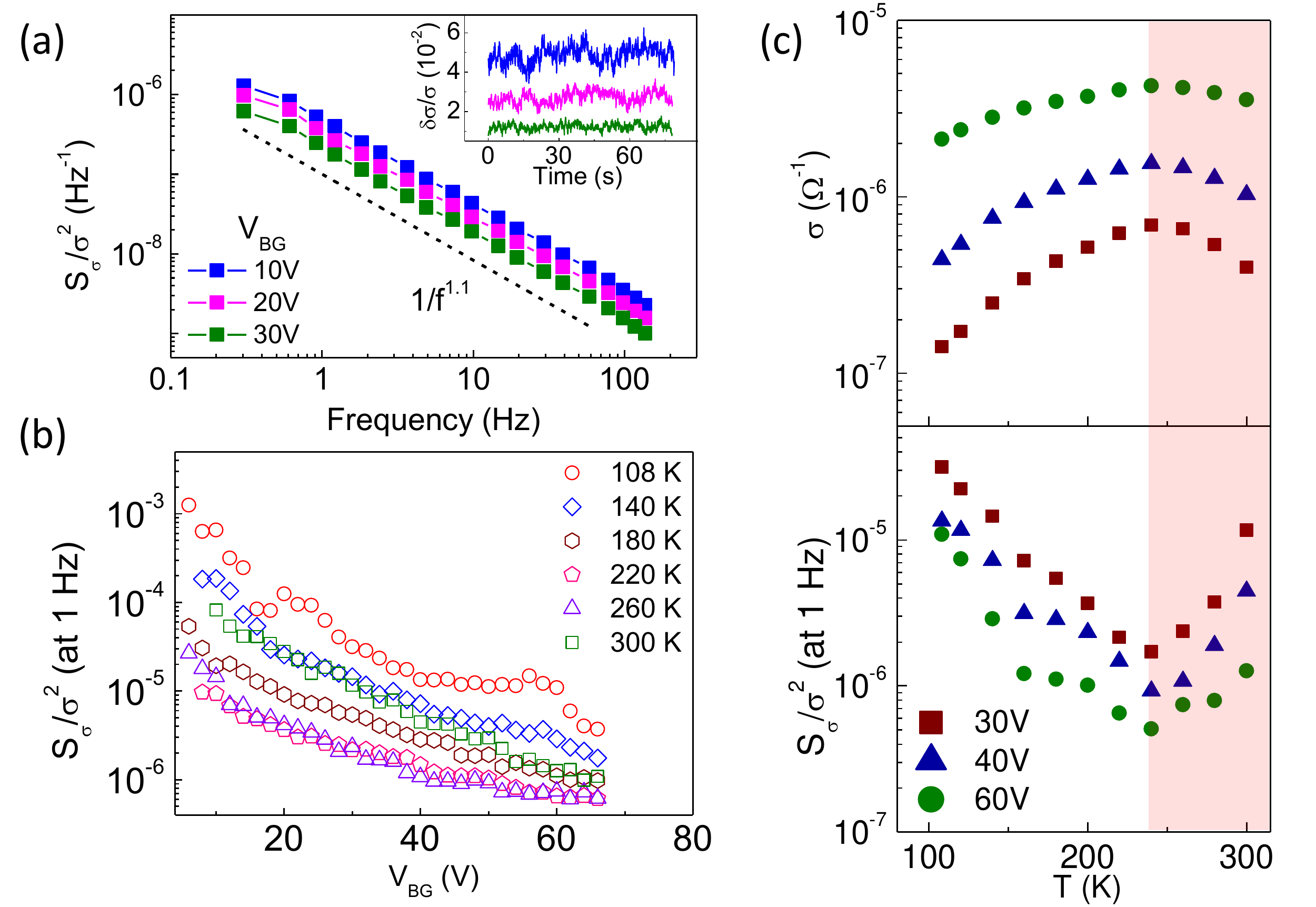}
\caption{(a)~1/$f$ noise power spectrum of conductivity fluctuations at three different $V_{BG}$. The inset shows corresponding conductivity
fluctuations as a function of time at the same gate voltages. (b)~Gate voltage dependence of noise power spectral density~$S_{\sigma}/\sigma^2$
at various $T$ for a single layer device. (c)~Non-monotonic $T$ dependence of both $\sigma$ and $S_{\sigma}/\sigma^2$ for 1L-2 device at three
different $V_{BG}$. The localized regime was observed below 240~K~(unshaded) and the weak metallic regime for $T~>$~240~K~(shaded).}
\end{center}
\end{figure*}

We now focus on the low-frequency conductivity fluctuations~(inset of Fig.~2a) in our MoS$_2$ devices. The normalized power spectral
densities~($S_{\sigma}/\sigma^2$) of conductivity fluctuations showed 1/$f$-type frequency spectrum at low frequencies~(Fig.~2a). The gate
voltage dependence of $S_{\sigma}/\sigma^2$ is shown in Fig.~2b. We observed that $S_{\sigma}/\sigma^2$ decreases monotonically with increasing
gate voltage at a fixed temperature. The temperature dependence of both $\sigma$ and $S_{\sigma}/\sigma^2$ at three fixed $V_{BG}$ are shown in
Fig.~2c. We observed that both change non-monotonically. The sharp decrease in $\sigma$ below 240~K can be readily attributed to the localized
state transport~\cite{natureofelectronic}, where $S_{\sigma}/\sigma^2$ increases exponentially due to the broad distribution of the waiting time
of the carriers between successive hops~\cite{shklovskii}. However, for $T~>~240$~K and especially at large $V_{BG}$, where $\sigma$ displays a
metal-like transport~\cite{metalinsulatormos2,pablomos2}, the noise magnitude increases with increasing temperature as in a diffusive
quasi-metallic systems~\cite{duttahorn}. It should be mentioned here that such weak diffusive transition was not observed in some devices till
300~K~(see supplementary material for more discussions).

\begin{figure*}
\begin{center}
\includegraphics[width=0.7\linewidth]{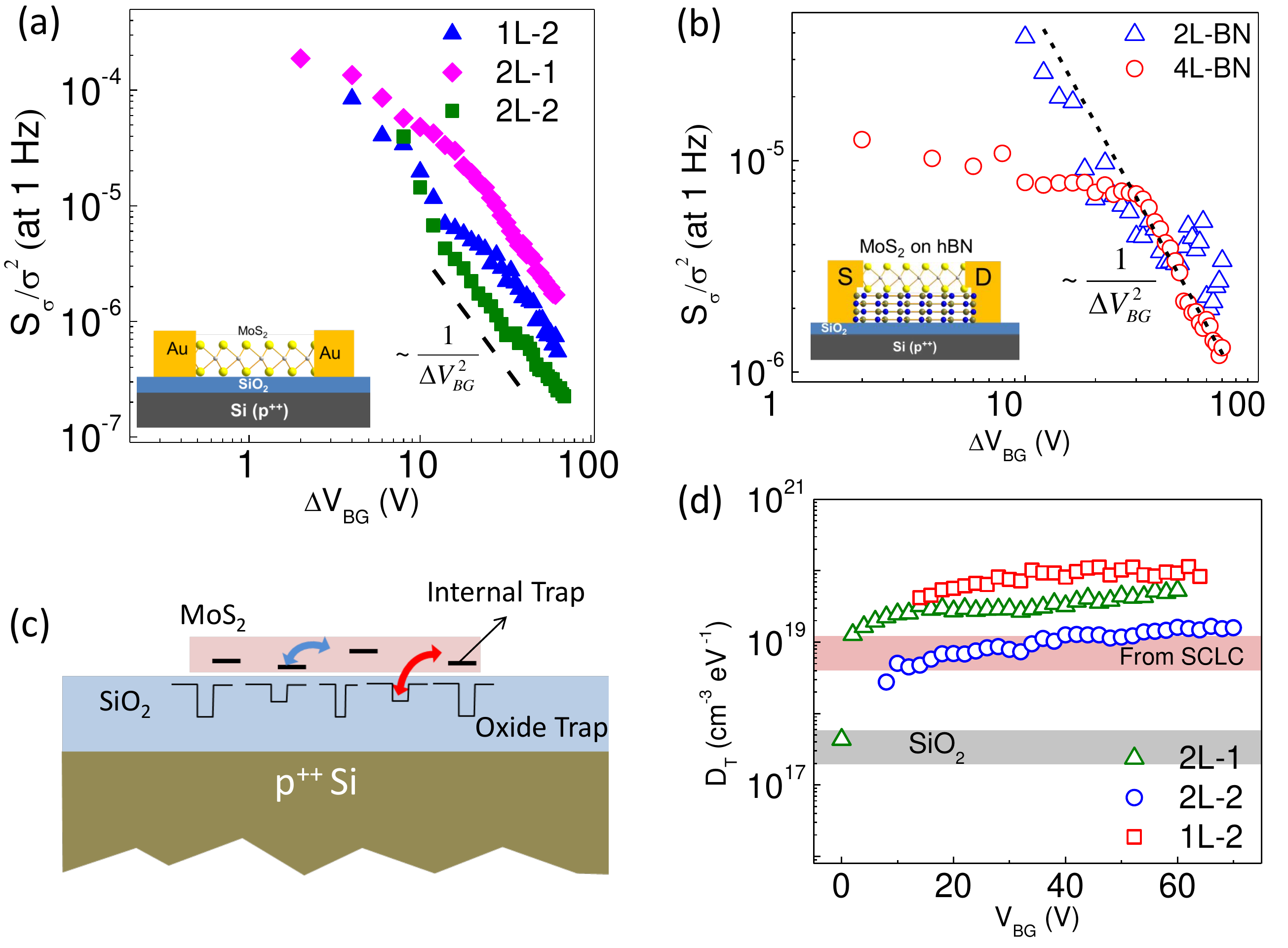}
\caption{(a)~$1/ \Delta V_{BG}^2$ dependence of $S_{\sigma}/\sigma^2$ at high $V_{BG}$ and near room temperature for three different MoS$_2$
devices on SiO$_2$ substarte. The schematic of an `MoS$_2$ on SiO$_2$' device is shown in the inset. (b)~Similar noise behaviour in `MoS$_2$ on
hBN' devices. The inset shows the schematic of a MoS$_2$ device on hBN substarte. (c)~Trap states in MoS$_2$ transistor can arise either due to
external oxide traps or internal structural disorders. (d)~Experimentally calculated trap density $D_T$ as a function of $V_{BG}$ for three
different devices. The same for standard SiO$_2$ substrate~(grey band) and from SCLC~(red band).
}
\end{center}
\end{figure*}

The carrier density dependence of $S_{\sigma}/\sigma^2$ for three different devices is shown in Fig.~3a. We observe that at low $\Delta V_{BG}$,
the variation differs from one sample to another and may be connected to the details of electron localization. However, at large $\Delta
V_{BG}$, {\it all} devices show $S_{\sigma}/\sigma^2 \propto 1/\Delta V_{BG}^2$, which is a characteristic feature of \emph{number fluctuation}
in semiconductors~\cite{mcwhorter}. Number fluctuation arises from trap states varying in energy and position within or close to the conduction
channel~(see schematic in Fig.~3c). The interfacial trap states in the SiO$_2$ substrate close to channel has been quantitatively shown to cause
similar scenario in graphene-on-SiO$_2$~\cite{ouracsnano,atindaprl,xudu} and Si-MOSFETs~\cite{jayaraman}. To check whether the oxide traps can
account for the observed noise, we calculate the oxide trap density $D_T$ per unit volume and energy using the trap-channel tunnel model,
developed originally for MOSFETs~\cite{loannidis,jayaraman}. The calculated $D_T$ has been plotted in Fig.~3d for three different devices which
is approximately two orders of magnitude higher than SiO$_2$ surface trap density~(see supplementary for detail calculations). To find an
alternative source, we calculate the trap density of states from its volume density $N_T$ obtained from the SCLC measurements~(Fig.~2d).
Assuming $N_T\approx 1~-~3\times 10^{17}$ ~cm$^{-3}$ as estimated from SCLC, and the trap states in an energy bandwidth of $k_BT\sim~26$~meV
contribute to carrier number fluctuation, we calculate trap state density $D_T~=~N_T/26\mbox{(meV)}\approx
4\times10^{18}~-~10^{19}$~cm$^{-3}$eV$^{-1}$ which reasonably agrees with the values calculated using number fluctuation model for devices 2L-1
and 2L-2~(see Fig.~3d). The higher value $D_T$ for 1L-2 device can be attributed to the error introduced during the calculation of
transconductance~$g_m$ due to contact resistance~(see supplementary material). A recent study has shown a similar order of trap density in
MoS$_2$ thin film transistors~\cite{balandinmos2noise}.
This establishes that a set of structurally originated Coulomb scatterer determine both electrical transport and low frequency noise in MoS$_2$
transistors whereas the interfacial oxide trap charges plays a minimal role in the transport properties.

 We further confirm this result by fabricating MoS$_2$ transistor on crystalline hexagonal boron nitride~(hBN) substrate known to be free from
surface trap states and dangling bonds~\cite{dean1}~(see inset of Fig.~3b and detail in supplementary). This device architecture eliminates the
oxide trap states to cause carrier exchange with the channel. Noise measurements on two `MoS$_2$ on hBN' devices are shown in Fig.~3b. We find
that $S_{\sigma}/\sigma^2 \propto 1/\Delta V_{BG}^2$ at large $\Delta V_{BG}$ which establishes number fluctuation as the dominant noise
mechanism providing further support to the internal origin of the localized trap states.

It is well-known that the presence of water vapour and fabrication-induced resist residues can also act as an additional source of trap states
in thin film transistors. In order to address the contribution of these external sources, we also fabricate devices by transferring a
thin~(20~nm) single crystalline layer of hBN on top of the MoS$_2$ flake prior to lithography processes~(see insets of Fig.~4b and 4d for
schematic and actual device images respectively). This protects the channel from acrylic residues and the possibility of presence of water
vapour is also minimal as the transfer was done at 100$^0$C. The measurements were performed before and after vacuum annealing at 150$^0$C and
4.6$\times$10$^{-6}$~mbar vacuum  for 3 hours. We found that in all cases the $V_{ON}$ shifted towards large ~($<$~-60~V) negative gate voltages
after annealing. This has been observed previously but the reason behind this remained controversial~\cite{pablomos2}. The transfer
characteristics for both kinds of devices are plotted in Fig.~4a and 4b respectively as a function of $\Delta V_{BG}$. We find that the
characteristics curves show similar density dependence beyond $\Delta V_{BG}~>~10$~V indicating a minimal role of the contacts in these devices.
The noise behaviour is shown in this regime for both kind of devices before and after annealing in Fig.~4c and 4d respectively. For comparison,
we define plot $S_I/I^2$ normalized to its magnitude at $\Delta V_{BG} \approx $10~V. We find that, in spite of having similar transfer
characteristics, the noise magnitude decreases by 10~-~30 times on annealing irrespective of the nature of surface protection. This observation
suggests that in our devices where the measurements are carried out in high vacuum condition, the noise is not dominated by the external sources
of trap states.

\begin{figure*}
\begin{center}
\includegraphics[width=0.8\linewidth]{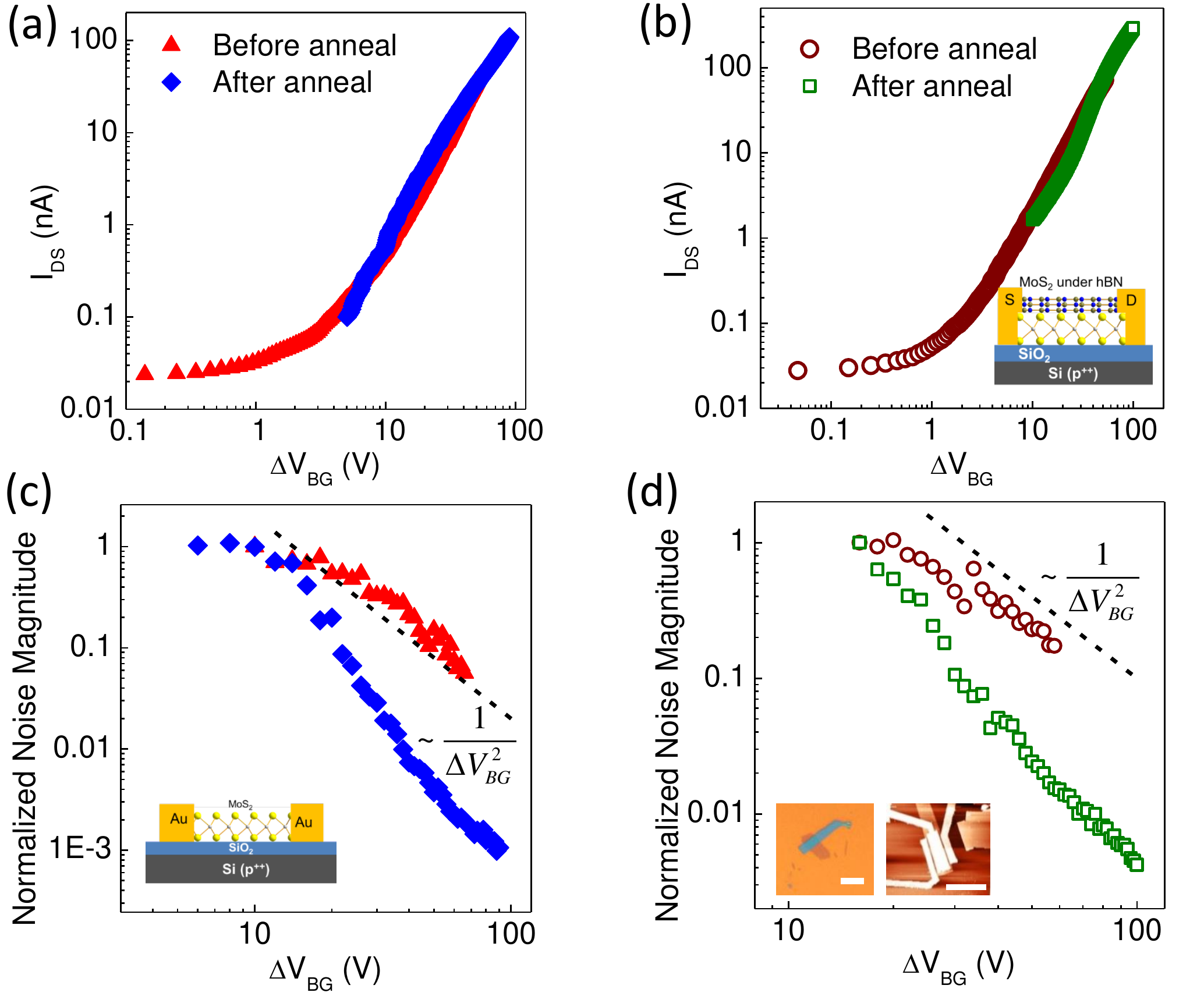}
\caption{(a),(b)~Transfer characteristics before and after annealing of MoS$_2$ channel with open and protected upper surface. The inset in 4b
shows schematic of an hBN protected MoS$_2$ device. (c),(d)~Noise behaviour in the same devices before and after annealing. $S_I/I^2$ was
normalized with its values at $\Delta V_{BG}~=~10~$V and 16~V in Fig.~4c and 4d respectively . The black dashed lines indicate $1/\Delta
V_{BG}^2$. The inset in figure 4d show an optical micrograph of hBN flake on single layer MoS$_2$ flake and atomic force microscopy image of
final device. Both the scale bars are 8~$\mu$m.}
\end{center}
\end{figure*}

The noise measurement in conjunction with the low temperature conductivity measurement can provide a crucial insight on the structural
morphology of the trap states. We observed that, at low temperature, conductivity as a function of gate voltage showed a number of reproducible
peaks even at high carrier density~($V_{BG}~\sim~$60~-~80~V~ and see Fig.~S7 in supplementary material and ref.~\cite{natureofelectronic}). This
indicates strong charge inhomogeneity in the MoS$_2$ channel. Oscillatory conductivity also suggests that the charge inhomogeneous
regions~(patches) act like quantum dots in semiconducting MoS$_2$ channel. These patches can exchange electrons with the channel to produce
1/$f$ noise~(Fig.~5a), and also act as charged impurities. For a more quantitative approach, we consider $D$ and $\Delta E$~=~$e^2/C$~( where
$C\approx4 \varepsilon_0 \varepsilon_r D$ is the capacitance) as the average diameter and charging energy respectively of the patches. Because
of their small size~($D \sim 10$~nm)~\cite{natureofelectronic}, we also considered single particle quantum level spacing as $\Delta E_Q$. The
scenario has been depicted in the schematic of Fig.~5a~(bottom). Assuming that the channel is nearly diffusive at high $T$ and noise arises due
to thermally activated exchange of charge carriers between channel and patches, the noise magnitude can be expressed as (see full derivation in
the supplementary material):

\begin{equation}
\label{eq2} \frac{\langle\delta \sigma^2\rangle}{\sigma^2} \approx \left[\frac{k_B T}{\Delta E_Q}\right] ^3 \frac{N_{E}}{N^2} \exp[-
\frac{\Delta E}{k_B T}]
\end{equation}

\noindent where $N_E$ and $N$ are the total number of patches and total number of electrons respectively in the channel. Equation~1 suggests an
exponential temperature dependence in diffusive regime~(shaded region in Fig.~2c) . To verify this, we chose the devices which showed an
insulator to weak diffusive transition~\cite{metalinsulatormos2,pablomos2} at high $T$. The variation of $ T^{-3} \langle\delta
\sigma^2\rangle/\sigma^2$ as a function of $1/T$~(see Fig.~5b) in the high $T$ region yields average charging energy $\Delta
E\sim$~88$~\pm$~15~meV for the device 1L-1. A similar calculation of $\Delta E$ for device 1L-2, which also showed a weak diffusive transition,
gives charging energy of 90$~\pm$~20~meV~(see supplementary Fig.~S8). It should be noted that charge exchange can last well upto the room
temperature and beyond due to large charging energy. We use the Hooge relation $S_{\sigma}/\sigma^2~=~\gamma_H/nAf$ to compare the noise
magnitude from different devices, where $\gamma_H$, $A$ and $n$ are the Hooge parameter, area of the channel, and electron density calculated
from $\Delta V_{BG}$ using the capacitance between the channel and the backgate~(The Hooge relation is not strictly valid in the McWhorter-type
number fluctuation noise, and should be taken only as a guideline). Since $\sigma \propto n^2$ (Fig.~1a) in the diffusive regime, $\gamma_H
\propto 1/\mu_{FE}$~(derivation in supplementary material and also see Ref[\cite{mos2noiseOhioGr}]), which provides a reasonable description of
the noise measured in this work (calculated at $\Delta V_{BG} \approx 58~V$, and $f = 1$~Hz), as well as that in
Ref[\cite{noisemos2hersem}]~(see Fig.~5c). We observe that the noise level is much higher in MoS$_2$ films compared to graphene device at
similar carrier densities and on same substrate. This also strengthens our conclusion of an internal origin of disorder in MoS$_2$ films.

\begin{figure*}
\begin{center}
\includegraphics[width=0.8\linewidth]{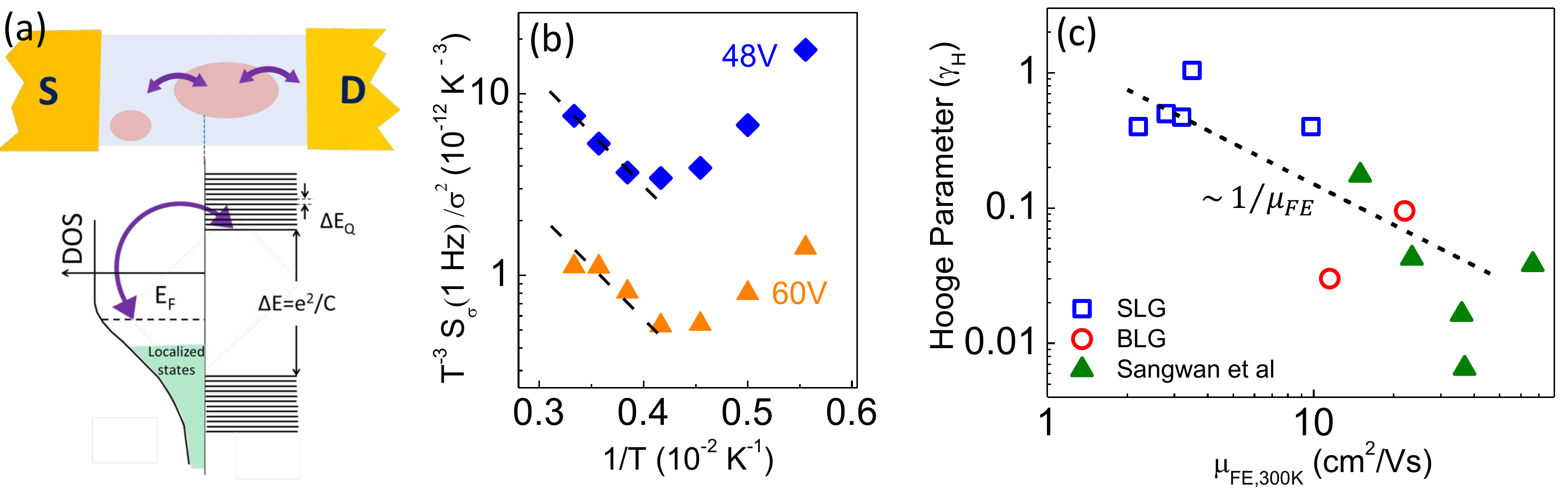}
\caption{(a)~Scematic representation of charge exchange between Coulomb blockaded localized state and semiconducting channel~(up). The same in
the energy diagram~(down). (b)~$T^{-3}S_{\sigma}/\sigma^2$ as function of $1/T$ for 1L-2 device at $V_{BG}~=~48, 60~V$. The slope of the
exponential fit gives the charging energy $\Delta E$. (c)~Hooge parameter~($\gamma_H$) as a function of field effect mobility~($\mu_{FE}$) for
five single layer~(blue square) and two bilayer~(red circle) devices at $\Delta V_{BG}\sim 58~V$ and 300~K. The filled triangles indicate the
$\gamma_H$ calculated in ref~[\cite{noisemos2hersem}] in similar MoS$_2$ devices. The dotted line represents variation of $\gamma_H$ inversely
proportional to $\mu_{FE}$.}
\end{center}
\end{figure*}

We now discuss the possible origin of these charge inhomogeneous patches in MoS$_2$ films. Recently, it has been reported that both $n$-type and
$p$-type charge inhomogeneity can occur in natural MoS$_2$ crystal, and are related to the crystal structure~\cite{defectmediated}. The low
sulphur concentration regions~(S-vacancies or Mo-like cluster) act as nanometer sized metallic regions and can take part in electron exchange.
Such sulphur deficit region in our devices can originate either during crystal growth or during mechanical exfoliation. Moreover, $p$-type
regions are formed due to structural defects to reduce strain in sulphur rich areas. We also discuss one more possibility which was revealed
from our XPS measurement where it was observed that the positions of the extra peaks match well with the metastable 1T phase of
MoS$_2$~\cite{photolumichemicallyexfo}. The 1T phase, which may nucleate around sulphur vacancies or other crystalline disorder, is metallic in
nature, and can introduce charge inhomogeneity in semiconducting 2H-MoS$_2$. The strong suppression in noise after annealing in both
surface-protected and unprotected devices provides additional support towards such bulk inhomogeneity-driven noise mechanism in MoS$_2$ FETs.
Annealing can not only modify the layout of individual defect or cluster, but also decrease the 1T fraction
significantly~\cite{photolumichemicallyexfo}. Finally, for a quantitative estimate, we calculate the number of charge inhomogeneous patches per
unit volume $N_p~\sim~0.05/D^2t~\approx~2\times10^{17}$cm$^{-3}$, where we take $D \sim 15~$nm, and $t\sim1$~nm as the thickness of MoS$_2$
layer. This value agrees closely with the $N_T$ value calculated from SCLC measurement. Moreover, we calculate $N_E~=~N_p\times At$, where $A$
is the area of the channel. Using equation~1 and some typical values of the parameters obtained from experiment, we obtain $\gamma_H \approx
0.2$ at $n~=~4.5 \times 10^{12} $~cm$^{-2}$~(see supplementary for calculation and equation~S15). This value agrees reasonably with the
$\gamma_H$ values obtained directly from the experiment~(See Fig.~4c).

In conclusion, we have studied electrical conductivity and low frequency noise in ultrathin MoS$_2$ transistor. We found that both are dominated
by same set of localized trap states. The trap density was calculated independently from both non-equilibrium conductivity and noise
measurements which agree well and turn out much higher that SiO$_2$ surface trap density. This strongly suggests that the trap states are not
external, but related to the crystal structure of the MoS$_2$ film that determines electron transport in MoS$_2$ field effect transistors.


\newpage

\begin{center}
{\bf \LARGE Supplementary information}
\end{center}

\begin{center} {\bf \Large Microscopic origin of charge impurity scattering and flicker noise in MoS$_2$ field effect transistors}
\end{center}

\subsection{Experimental details:} For `MoS$_2$ on SiO$_2$' devices, MoS${_2}$ flakes were exfoliated on SiO$_2$
(300~nm)/n$^{++}$ Si wafer from bulk MoS$_2$ crystals (SPI Supplies) using scotch tape. To keep the disorder level comparable, the wafers were
thoroughly cleaned by standard RCA cleaning followed by acetone and isopropyl alcohol cleaning in ultrasonic bath. The flakes were identified
initially by optical microscope~\cite{visibilitykis,visibilitygomez}. The thickness and quality of each flake was determined by Raman
spectroscopic measurement~\cite{Ramanacsnano}(see Figure~S1). The Raman data was recorded using WITEC confocal (X100 objective) spectrometer
with 600 lines/mm grating, 514.5 nm excitation at a very low laser power level (less than 1 mW) to avoid any heating effect. Au(40 nm) contacts
were defined using standard electron beam lithography followed by thermal evaporation of 40~nm Au and lift-off in hot acetone. No Ar/H$_2$
annealing was done in any of our devices after liftoff because we found change in morphology of only Au (no underlayer like Ti or Cr) pads on
SiO$_2$ substrate after annealing beyond 250$^o$C.

To transfer MoS$_2$ on hexagonal boron nitride~(hBN: from Momentive), we prepared two different substrates~\cite{weesgroup}. First, 10-20 nm hBN
on Si/SiO$_2$ wafer and a glass slide coated with transparent tape and 400 nm thick EL9 (Microchem). MoS$_2$ flakes were exfoliated using scotch
tape technique on the glass-tape-EL9 stack. The transfer was done in MJB3 Mask aligner with a heated stage at 100$^0$C to reduce the water
vapour at the interface~(see Fig.~S5). The MoS$_2$ devices with protected upper surface of the channel were fabricated in the similar process as
discussed previously. But the exfoliation was done in the reverse way.
 \begin{figure}[h]
 \centering
 \includegraphics[width=0.9\textwidth]{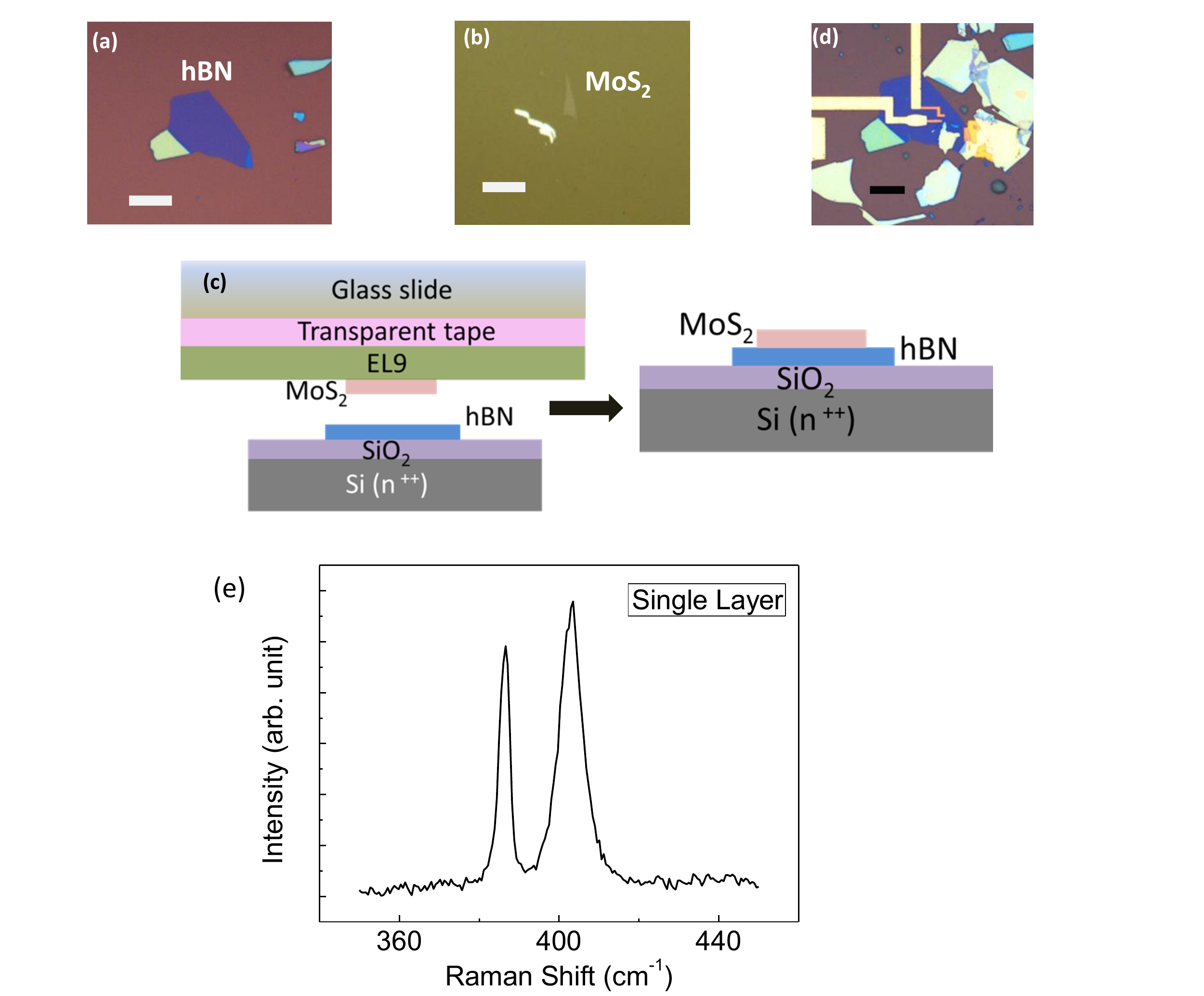}
 \caption{(a)~Thin layer of hBN on 275 nm of Si/SiO$_2$ substrate. (b)~Single layer MoS$_2$ flake on glass-tape-EL9 stack. (c)~Transfer of MoS$_2$
 flake from polymer onto hBN. (d)~Final MoS$_2$ on hBN device after liftoff. (e)~Raman spectrum of a single layer MoS$_2$ film on Si/SiO$_2$ substrate
 after exfoliation.
 }
 \end{figure}

The devices were wire-bonded in ceramic leadless chip carrier obtained from Kyosera. The devices were heated till 120$^0$C on a hotplate and
immediately transferred into the low temperature cryostat. The cryostat was evacuated till 2$\times$10$^{-6}$~mbar pressure before starting
experiments to reduce water vapour from the surface of the MoS$_2$ channel. All the measurements were carried out in same ultrahigh vacuum
condition.

\subsection{Details of devices measured:}
The detail of the devices are given in Table~I.
\begin{table}[h]
\caption{\label{table1}Details of the devices:}
\begin{ruledtabular}
\begin{tabular}{cccc}
 Device &  Layer Number & Device area (L$\times$W) \footnote{both dimensions in $\mu$m} & Mobility($\mu_{FE}$)\footnote{in cm$^2$/V-s near room temperature and $\Delta V_{BG}\sim60V$}\\ \hline
 1L-1 & 1 & $2 \times 2.5$ & 3  \\
 1L-2 & 1 & $5 \times 8$ & 9 \\
 1L-3 & 1 & $ 0.6\times 4.9$ & 2.2 \\
 1L-4 & 1 & $ 1.1\times 4.9$ & 2.8 \\
 1L-5 & 1 & $ 2.1\times 4.8$ & 3.2 \\
 1L-6 & 1 & $ 2.7\times 1.6$ & 0.7 \\
 1L-7 & 1 & $ 2\times 4$ & 10 \\
 1L-8 & 1 & $ 0.16\times 1.9$ & 0.5\\
 1L-9 & 1 & $0.8 \times 4 $ & 2.6 \\
 2L-1 & 2 & $2.8 \times 2.5$ & 22 \\
 2L-2 & 2 & $3 \times 4.9$ & 11 \\
 2L-3 & 2 & $1.9 \times 2.2$ & 10 \\
1L-BN & 1 & $3 \times 4 $ & 4 \\
2L-BN & 2 & $0.8 \times 1.8$ & 3 \\
4L-BN & 4 & $1.7 \times 2.7$ & 20 \\

\end{tabular}
\end{ruledtabular}
\end{table}

\subsection{Noise measurement Scheme:}

For noise measurement the sample was biased with a constant ac voltage $\sim10~$mV at 226 Hz from lockin. The sample current was passed through
a low noise preamplifier and measured using lockin technique. The current fluctuation data as a function of time, called time series, was
acquired with a high sampling rate data card. The time series data was Fourier transformed to obtain current noise power spectral density
$S_I/I^2$ as a function of frequency $f$~\cite{scofield1byfnoise,arindamdetailnoisearxiv}. The current power spectral density can be converted
to conductivity fluctuation power spectral density $S_{\sigma}/\sigma^2$ using the relation
\begin{equation}
S_{\sigma}/\sigma^2=S_I/I^2.
\end{equation}
In the figures of the main manuscript and supplementary, we have plotted either ``$S_{\sigma}/\sigma^2$ at 1~Hz'' or ``integrated noise power
$\langle\delta\sigma^2\rangle / \sigma^2$'' as a measure of noise. $\langle\delta\sigma^2\rangle / \sigma^2$ is defined as the power spectral
density integrated over the measurement frequency bandwidth i.e.
\begin{equation}
\langle\delta\sigma^2\rangle / \sigma^2=\int_{f_1}^{f_2} S_{\sigma}/\sigma^2 df
\end{equation}
where $f_1$ and $f_2$ are the lower and upper cut-off frequencies during the measurement.

The relation between $\frac{\langle\delta \sigma^2\rangle}{\sigma^2}$ and $\frac{S_{\sigma}}{\sigma^2}$ at 1~Hz can be expressed as shown below:
\begin{eqnarray*}
\frac{\langle\delta \sigma^2\rangle}{\sigma^2} &=& \int_{f_1}^{f_2} \frac{S_{\sigma}}{\sigma^2} df\\
\end{eqnarray*}
and from Hooge relation, we have
\begin{equation}
\frac{S_{\sigma}}{\sigma^2}=\gamma_H/nAf^{\alpha}
\end{equation}
Therefore, $|\frac{S_{\sigma}}{\sigma^2}|_{1~Hz}~=~\gamma_H/nA$ and
$\frac{S_{\sigma}}{\sigma^2}~=~\frac{|\frac{S_{\sigma}}{\sigma^2}|_{1~Hz}}{f}$

For $1/f$ noise $\alpha \sim 1$ and hence
\begin{eqnarray*}
\frac{\langle\delta \sigma^2\rangle}{\sigma^2} &=& \int_{f_1}^{f_2} \frac{S_{\sigma}}{\sigma^2} df\\
                                               &=& \int_{f_1}^{f_2} \frac{|\frac{S_{\sigma}}{\sigma^2}|_{1~Hz}}{f} df\\
                                               &=& |\frac{S_{\sigma}}{\sigma^2}|_{1~Hz} \int_{f_1}^{f_2} df/f\\
                                               &=& |\frac{S_{\sigma}}{\sigma^2}|_{1~Hz}~ ln(f_2/f_1)\\
\end{eqnarray*}

Therefore $\frac{\langle\delta \sigma^2\rangle}{\sigma^2}$ and $|\frac{S_{\sigma}}{\sigma^2}|_{1~Hz}$ are proportional to each other by a
constant factor.


\subsection{XPS measurement and data processing scheme:}

The X-Ray Photoelectron Spectroscopy~(XPS) measurements were performed on bulk MoS$_2$ with a commercial electron spectrometer from VSW
Scientific instrument at a base pressure of $5\times 10^{-10}$~mbar. Mo $3d$ and S $2p$ core level spectra were recorded with Al K$\alpha$
radiation (photon energy 1486.6 eV) at a pass energy of 20 eV. The core level spectra were corrected for background using the Shirley algorithm,
and chemically distinct species were resolved using nonlinear least-squares fitting procedure. A Lorentzian function representing the lifetime
effect, convoluted with a Gaussian function representing the resolution was used to simulate the XPS peak shape. In order to minimize the number
of free parameters in the decomposition process, we impose several constrains like, the same spin orbit splitting of various component feature
of a particular core levels and the well-known branching ratios between the two spin-orbit split components.

\subsection{XPS spectrum obtained from second time measurement:}

We perform XPS measurements on two freshly cleaved samples. Results obtained from the first samples are shown in Fig. 1a and 1b of the main
manuscript. The intense pair of peaks (blue solid line) at 229.6 eV and  232.8 eV binding energies (BE) with a separation of 3.2 eV, can be
easily associated with the spin orbit split 3$d_{5/2}$ and 3$d_{3/2}$ pair of Mo$^{4+}$ of  2H phase of MoS$_2$~\cite{photolumichemicallyexfo}.
S 2p spectra in Fig 1b also has the main contribution from a spin orbit pair of the 2H phase of MoS$_2$ (blue solid line). In order to simulate
the experimental spectra properly, we need to incorporate extra features (red solid line) in both S 2p and Mo 3d spectra.  Relative intensities
of these features, plotted in red in Figs 1a and 1b compare to the intensities of the features drawn in blue appear quite the same (~5\%
relative intensity) in all figures, providing a consistency check from the independent analysis of Mo 3d and S 2p spectral features. Since the
intensity of these extra species were very small compared to the main peaks, we performed the same XPS experiment in another freshly cleaved
bulk crystal~(see Fig.~S2). The number of components and their relative intensity ratios are found to be the same for this sample as compared to
the one reported in the main text. Additionally the small peak at higher BE side of the main Mo$^{4+}$ feature in Mo 3d spectra probably coming
from small amount of oxidized Mo6+ present in this sample as the BE of this feature matches well with what is reported in the literature for
Mo$^{6+}$ ions~\cite{photolumichemicallyexfo,DDxpsPRL,DDandCNRRao}.

 \begin{figure}[h]
 \centering
 \includegraphics[width=0.9\textwidth]{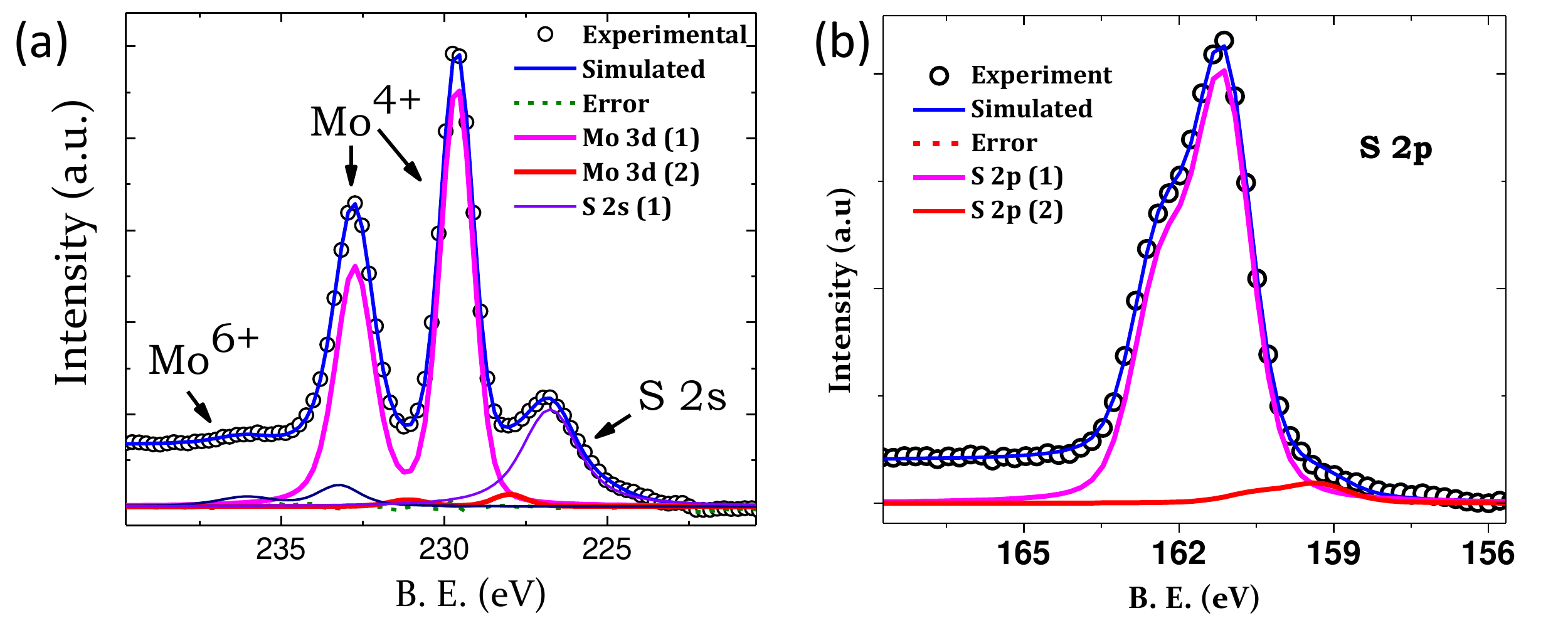}
 \caption{XPS spectra of (a)~Mo3d, S2s and (b)~S2p taken from freshly cleaved bulk MoS$_2$ crystal at room temperature.}
 \end{figure}

\subsection{n-type doping in MoS$_2$ :}
The origin of such n-doping is not very clear. Although, it was suggested that presence of halogen~(Cl, Br) impurity in crystal may lead to
n-type doping~\cite{Yin}, we couldn't find presence of halogen impurity in bulk natural crystal by EDX and XPS studies~(see supplementary).
Later on, it was also proposed that n-doping can possibly come from S vacancies in the crystal~\cite{superconductingdome} but a recent $T$
dependent study reveals that this may not be the dominant cause~\cite{pablomos2}.

We perform EDX and XPS study~(see Fig.~S3 and inset respectively) and find no presence of halogen impurity in naturally occurring MoS$_2$
crystal as discussed in Ref.~\cite{Yin}.
\begin{figure}
 \centering
 \includegraphics[width=0.6\textwidth]{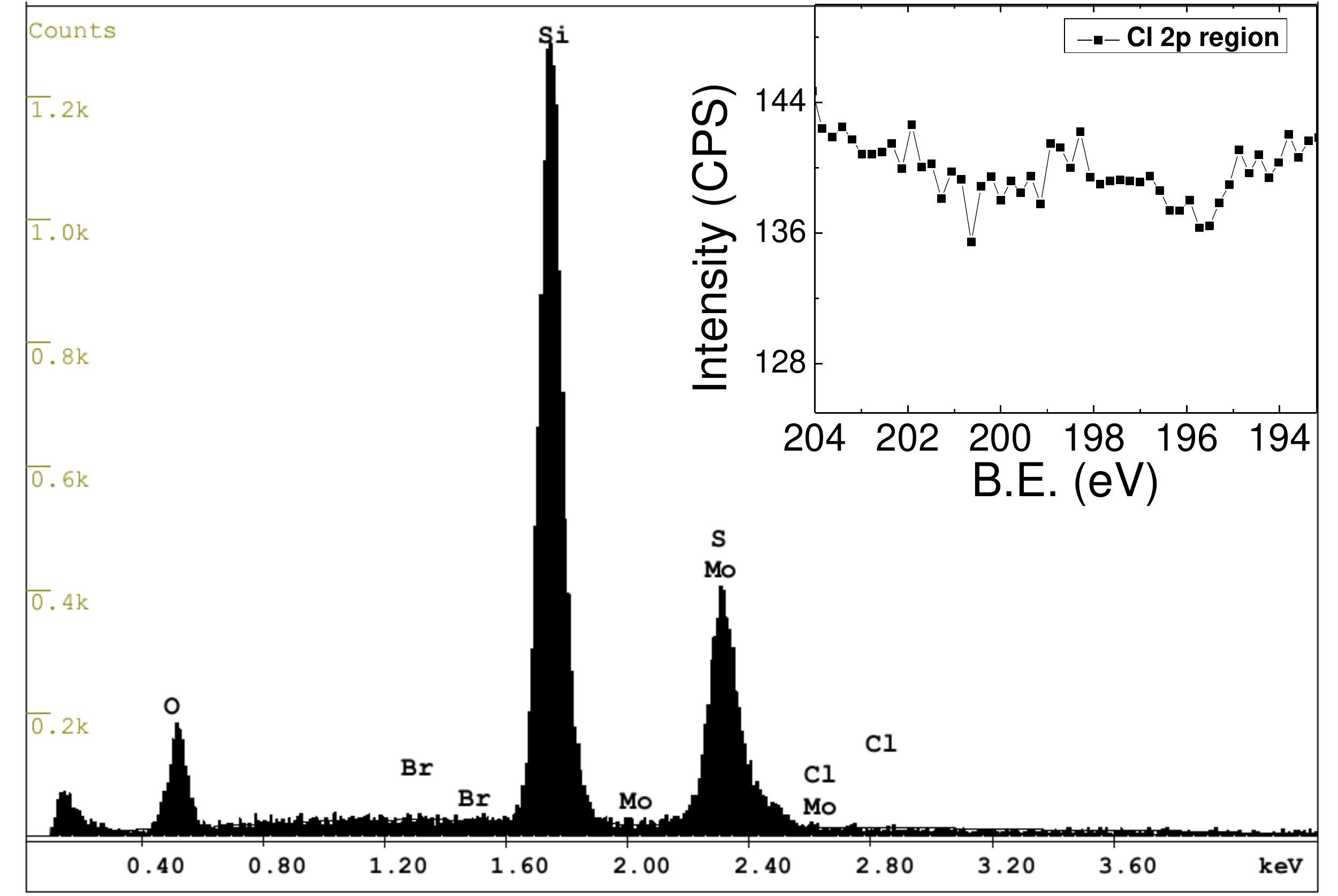}
 \caption{EDX spectrum of bulk MoS$_2$ crystal taken at room temperature and high vacuum condition. Inset shows XPS scan of chlorine 2p region. }
 \end{figure}

\subsection{Localized to weak diffusive transition:}

We believe that the localized electronic states in MoS$_2$ transistor become nearly extended with increasing carrier density~($n$) and $T$. This
manifests as a localized to weak diffusive transition in the system as shown in Fig.~S4(a). It is evident from the figure that transition
temperature changes with carrier density as transfer characteristics at 200K and 300K intersects at $V_{BG}$=54V, whereas same for 240K, 260K
and 300K intersects at $V_{BG}$=30V. Similar observations has been reported recently by two other groups~\cite{metalinsulatormos2,pablomos2}.
Here we mention that we didn't see this weak diffusive transition in every device till room temperature and $V_{BG}$ as high as $ \sim$ 70V. We
believe occurrence of such transition probably depends on intrinsic disorder landscape of individual flake. There the $\sigma$ vs. $T$ will look
like the unshaded region of Fig.~2c in the main text or as shown in Fig.~S4(b).

\begin{figure}
 \centering
 \includegraphics[width=0.7\textwidth]{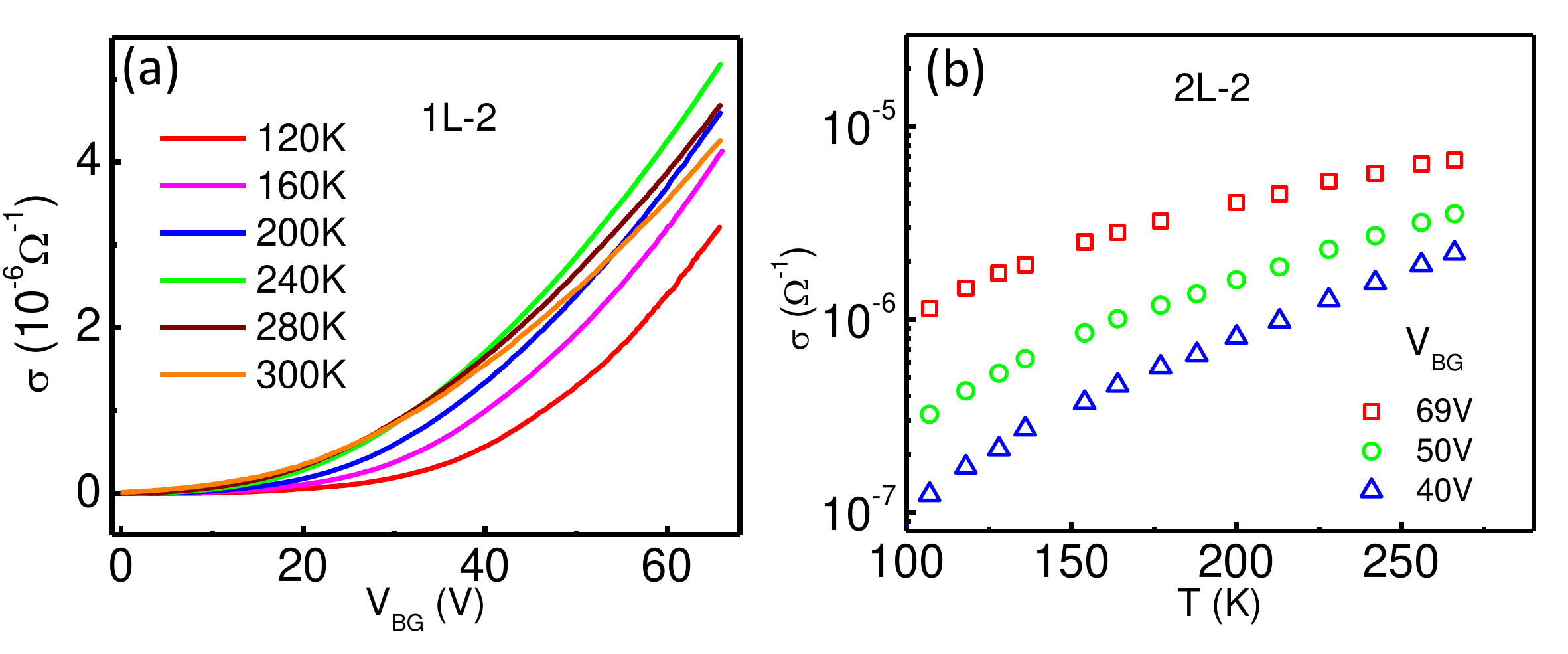}
 \caption{(a)~Backgate transfer characteristics of 1L-2 device showing localized to a weak diffusive transition at high $T$ and $V_{BG}$. (b)~$\sigma$
 as function of $T$ at three different $V_{BG}$ in a MoS$_2$ device with no localized to weak diffusive transition till room temperature. }
 \end{figure}

\subsection{Current-Voltage characteristics at 300K:}
The typical current voltage relationship obtained in our devices near room temperature and low temperature are shown in Fig.~S5. Such symmetric
and linear current-voltage characteristics excludes dominance of Schottky barrier at the contacts.

\begin{figure}
 \centering
 \includegraphics[width=0.7\textwidth]{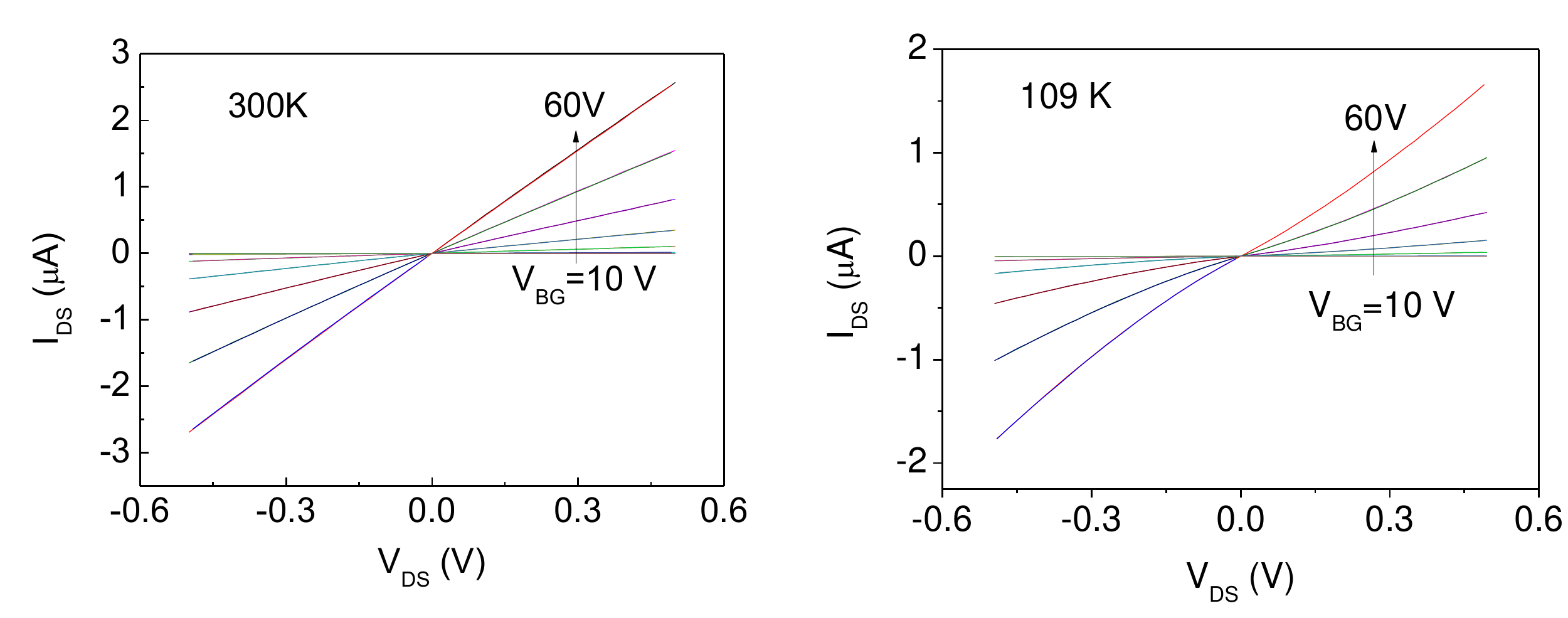}
 \caption{Current-voltage characteristics of a typical single layer device at 300~K and 109~K at different bacgate voltages.}
 \end{figure}

\subsection{Noise measurement in ultrathin MoS$_2$ devices:}
Before the noise measurement, all the devices were checked with $I^2$ dependence of variance of noise $S_I$ at low bias current to avoid any
heating induced effect~(see Fig.~S6a). The noise measurements were carried out at fixed $T$ for different gate voltages along the transfer
characteristic curve. We found that although the integrated noise power $\langle\delta\sigma^2\rangle / \sigma^2$ monotonically decrease with
increasing $V_{BG}$, the $T$ dependence is non-monotonic as shown in Fig.~S3b for device 1L-2. Such a non-monotonic $T$ dependence also
eliminate possibility of dominant contact noise in our devices. As we have already mentioned that all the devices have not shown such
transition. In those cases noise monotonically decreases till room temperature as shown Fig.~S6c.

\begin{figure}
 \centering
 \includegraphics[width=0.7\textwidth]{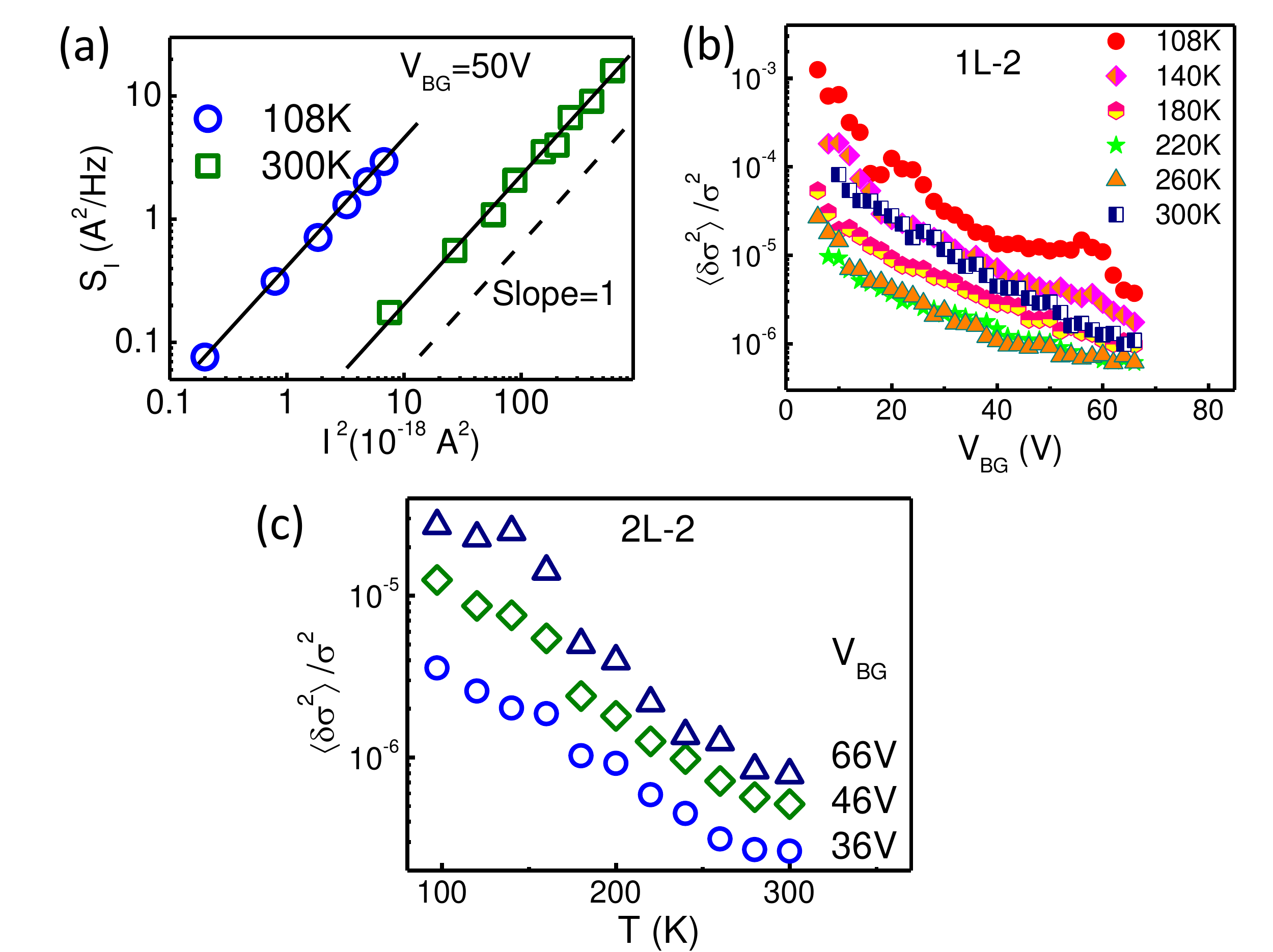}
 \caption{~(a)~$I^2$ dependence of variance of noise $S_I$ at high $V_{BG}$ and low bias current showing a linear behavior both at high and low $T$.
(b)~Gate voltage dependence of normalized $1/f$ noise power integrated over the measurement bandwidth at various $T$ for a single layer device.
We found noise magnitude monotonically decrease as $V_{BG}$ increases.(d)~Normalized conductivity noise~($\langle\delta\sigma^2\rangle /
\sigma^2$) shows a monotonic decrease as $T$ increases in a MoS$_2$ device with no localized to weak diffusive transition.}
 \end{figure}

\subsection{Trap-channel tunnel model for pure number fluctuation:}

For pure number fluctuation, the source-drain current noise can be written as~\cite{loannidis,jayaraman}
\begin{equation}
\frac{S_I}{I^2}=\left (\frac{g_m}{I^2} \right)^2 \frac{q^2k_BTD_T}{WLC_{ox}^2f\alpha}
\end{equation}
where $g_m$ is device transconductance, $q$ is electronic chrge, $k_BT$ is the thermal energy, $D_T$ is trap density at channel-substrate
interface per unit volume and energy, $WL$ is active device area, $C_{ox}$ is gate oxide capacitance per unit area, $f$ is the frequency and
$\alpha$ is the tunnelling attenuation coefficient of electronic wavefunction in SiO$_2$ $\sim 10^{10}m^{-1}$.

 The noise measurement was performed at different $V_{BG}$ with 2V interval along the transfer characteristic curve. Therefore, at a fixed gate voltage the
 current through the channel is constant.  Then we can write equation S4 as
\begin{equation}
S_I=\frac{g_m^2q^2k_BTD_T}{WLC_{ox}^2f\alpha}.
\end{equation}

Using equation~S5, We calculated $D_T\approx 6\times10^{19}-7\times10^{20}$ cm$^{-3}$eV$^{-1}$ in all our devices at room temperature. It has
been already discussed in literature that due to contact resistance $g_m$ is underestimated from transfer characteristic curve by 3-5
times~\cite{kisintegratedcircuit,intrinsicajayan,mobilitydualgatefuhrer,mobilitydualgatekis}. Therefore considering $g_m^2$ overestimates $D_T$
by a factor of 10, the corrected value of $D_T\approx6\times10^{18}-7\times10^{19}$ cm$^{-3}$eV$^{-1}$.

The surface trap charge density of SiO$_2$ is known to be $D_T~\approx~5\times 10^{17}$cm$^{-3}$ eV$^{-1}$ near room
temperature~\cite{jayaraman} which is one to two order less than the value obtained from our experiment.


\subsection{Oscillatory conductance at low temperature: } We observed reproducible oscillations in conductivity due to resonant tunnelling
between localized sites as shown in Fig.~S7~(left)~\cite{natureofelectronic}. It was found that the peaks in oscillation shift as a function of
both source-drain and gate bias as observed in quantum dot. A similar measurement is shown for device 1L-7 in Fig.~S7~(right) at 18~K where we
plot differential conductance~$dI/dV$ as a function of $V_{DS}$ and $V_{BG}$. The shifting of conductance peaks in ($V_{DS}, V_{BG}$) plane is
shown by the white lines which indicates two important consequences: First, the localized sites are not single particle localized states with
large distribution of charging energies, and second, approximate charging energy of the localized sites are $\sim$~80~-~100~meV.
 \begin{figure}
 \centering
 \includegraphics[width=0.7\textwidth]{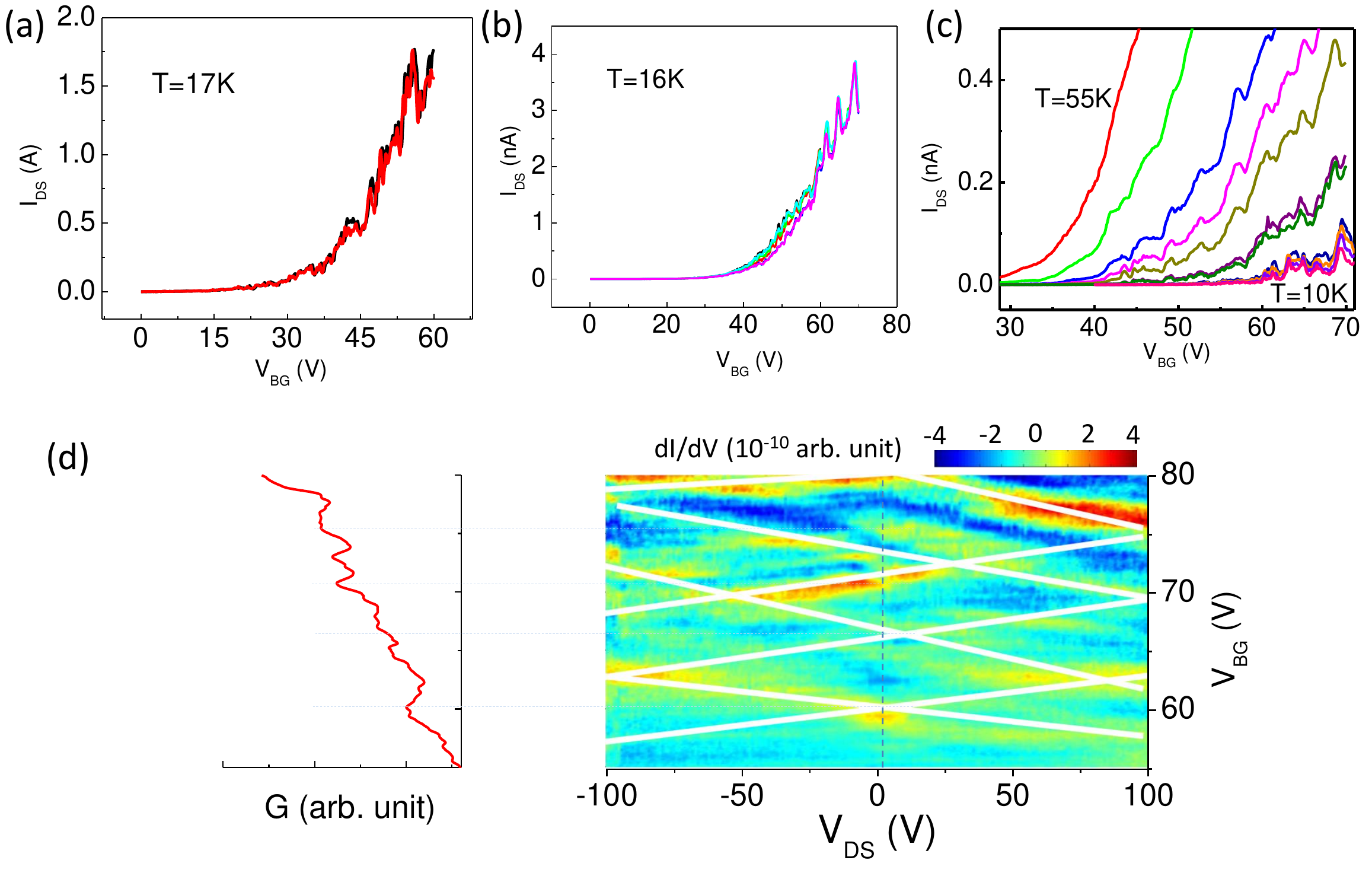}
 \caption{(a)-(c)~Reproducible oscillations in conductivity at low temperature for three different devices. (d)~2D map of differential
 conductance $dI/dV$ as a
function of $V_{DS}$ and $V_{BG}$ at $T~=$~18~K. The white lines shows the evolution of the resonant tunnelling peaks in ($V_{DS}, V_{BG}$)
plane~(right). Oscillations along the dashed line (left).}
 \end{figure}

\subsection{calculation of $\Delta E$ for 1L-2 device:} This device also showed a weak diffusive transition but the $T$ dependence was rather
weak. Hence for this device we plot $\left <\delta \sigma^2 \right >/\sigma^2$ as a function of $1/T$ which gives charging energy of
90$\pm$20~meV, which is in close agreement with the value of $\Delta E$, obtained from resonant tunnelling experiments. The number of data
points are limited because we exclude the data beyond 300~K due to onset of hysteresis. In Fig.~S8, the exponential fit yields, $\Delta E/k_BT
\approx 1033 \Rightarrow \Delta E=1033k_B/e$=~90~meV.

\begin{figure}
 \centering
 \includegraphics[width=0.5\textwidth]{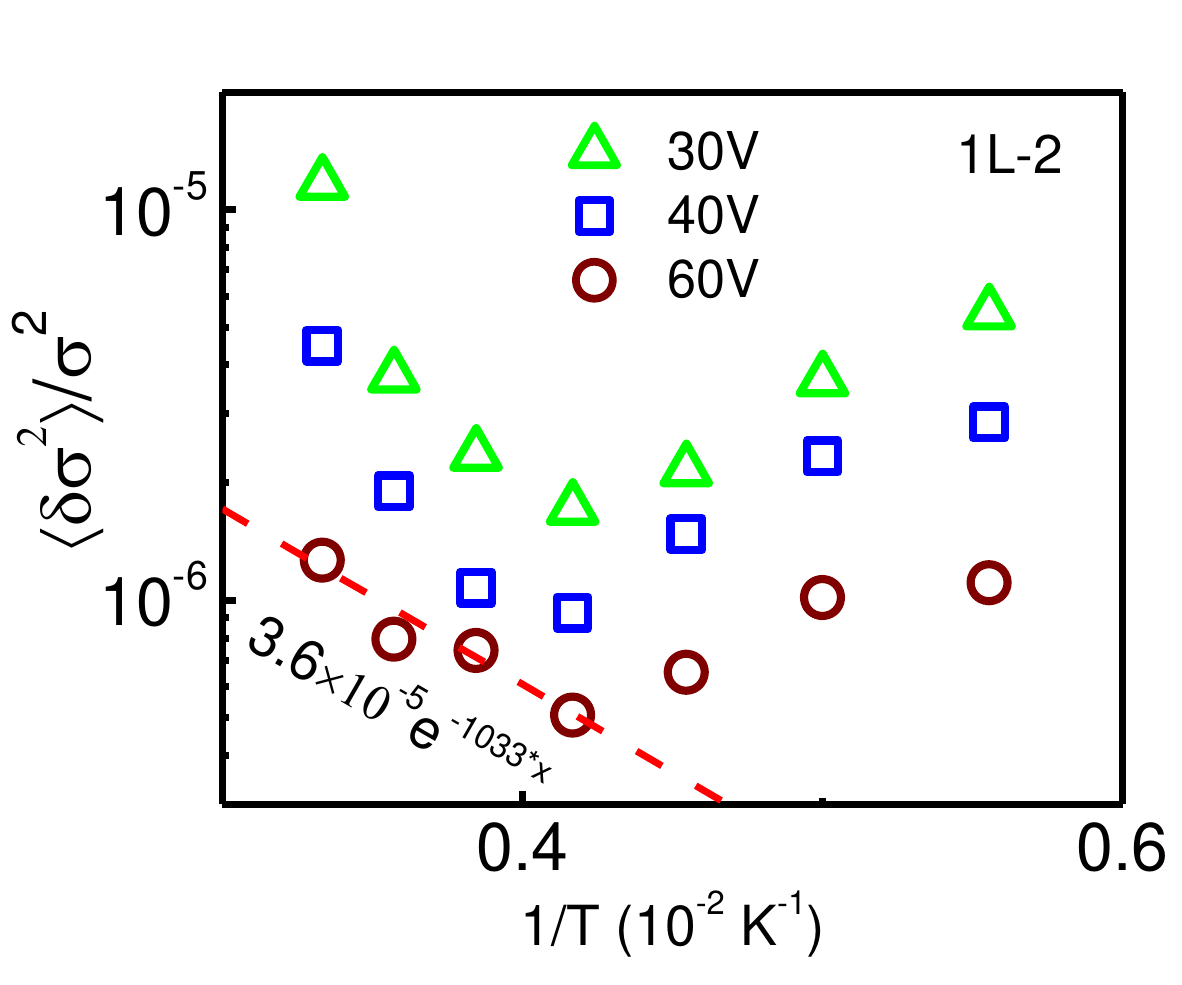}
 \caption{(a)~Normalised noise power $\langle\delta\sigma^2\rangle/\sigma^2$ as a function of $1/T$.}
 \end{figure}

\subsection{Carrier density dependence of Hooge parameter~($\gamma_H$):}
For 1/$f$ noise, the Hooge relation can be written as $S_{\sigma}/\sigma^2~=~\gamma_H/nAf$. In the diffusive regime, $S_{\sigma}/\sigma^2
\propto 1/n^2$ and $\sigma \propto n^2$ i.e. $\mu_{FE} \propto n$. Therefore, we have
\begin{equation}
\gamma_H \propto \frac{1}{\mu_{FE}}
\end{equation}

\subsection{Derivation of gate voltage and temperature dependence of noise:}
At high $T$ and $V_{BG}$, the localized states in 2H phase becomes nearly extended, and we assume $\sigma$~=~$ne$$\mu$ is valid. In this regime,
we get
\begin{equation}
\frac{\delta \sigma}{\sigma}=\frac{\delta n}{n}+\frac{\delta \mu}{\mu}
\end{equation}

\noindent where $\delta n$ is the number of the carrier per unit area being exchanged between the semiconducting 2H phase and electron-dense
patches. $\delta \mu$ is the mobility fluctuation term. As number fluctuation is the dominant source of $1/f$ noise, we assume contribution from
the second term is small compared to first term. Therefore we have
\begin{equation}
\frac{\delta \sigma}{\sigma}=\frac{\delta n}{n} \Rightarrow \frac{\langle\delta \sigma^2\rangle}{\sigma^2}=\frac{\langle\delta
n^2\rangle}{n^2}=\frac{\langle\delta N^2\rangle}{N^2}
\end{equation}
\noindent where we have multiplied both the denominator and numerator by the area of 2H phase $A_{2H}$ which is almost similar to total active
device area $A$ as 2H is the major phase in MoS$_2$. $N$ is the total number of the carriers in the 2H phase at a certain $V_{BG}$.
$\langle\delta N^2\rangle$ is the variance in total number of carriers fluctuating between the channel and the localized sites. The Fermi level
$E_F$ in the 2H phase will reach close to the mobility edge at high $V_{BG}$ as shown in Fig.~4a~(main manuscript). On the other hand, the
nanometer-sized electron-dense regions will act as quantum dot and the highest filled and lowest empty energy levels will be Coulomb blockaded
by the charging energy $\Delta E=e^2/C$, where $C$ is the average capacitance of an localized region with respect to surroundings.

As the number fluctuation is happening predominantly between the channel and the electron-dense patches then $\langle\delta N^2\rangle$ in
2H-channel will be related to $\langle\delta N_{E}^2\rangle$ in the patches by the following equation
\begin{equation}
\langle\delta N^2\rangle=N_{E} \langle\delta N_{E}^2\rangle
\end{equation}
where $N_{E}$ is the total number of nanometer-sized electron-dense patches in the device area and $\langle\delta N_{E}^2\rangle$ is the
occupation fluctuation of a single electron-dense patch.

We will now focus on a single electron-dense patch. We define the probability that there will be no electron inside a patch as $p_0$. Similarly,
probability of one electron as $p_1$ and for n electrons as $p_n$. Therefore, we have
\begin{eqnarray*}
p_0 + p_1 + p_2 + p_3 + p_4 +..............= 1\\
\Rightarrow p_0 + p_0 \exp[-\frac{\Delta E}{k_B T}] + p_0 \exp[-\frac{\Delta E + \Delta E_Q}{k_B T}] + p_0 \exp[-\frac{\Delta E + 2 \Delta
E_Q}{k_B T}] + .........= 1\\
\Rightarrow p_0 \{ 1+ \exp[-\frac{\Delta E}{k_B T}] + \exp[-\frac{\Delta E + \Delta E_Q}{k_B T}] + \exp[-\frac{\Delta E + 2 \Delta E_Q}{k_B T}]
+ ...........\} = 1\\
\Rightarrow p_0 \{ 1+ \exp[-\frac{\Delta E}{k_B T}]( 1 + \exp[-\frac{ \Delta E_Q}{k_B T}] + \exp[-\frac{ 2 \Delta E_Q}{k_B T}]
+ ...........)\} = 1\\
\Rightarrow p_0 \{ 1+ \exp[-\frac{\Delta E}{k_B T}] \times \frac{1}{1-\exp[-\frac{ \Delta E_Q}{k_B T}]} \} = 1\\
\Rightarrow p_0 \{ 1+ \exp[-\frac{\Delta E}{k_B T}] \times \frac{1}{\epsilon} \} = 1 \hspace{10mm}where~ \epsilon = 1-\exp[-\frac{ \Delta E_Q}{k_B T}] \\
\end{eqnarray*}
\begin{equation}
\Rightarrow p_0 = \frac{\epsilon}{\epsilon + \exp[-\frac{\Delta E}{k_B T}]}
\end{equation}

In order to evaluate $\langle\delta N^2\rangle$~(equation S9), we first calculate $\langle\delta N_{E}^2\rangle$ using
\begin{equation}
\langle\delta N_{E}^2\rangle = \langle N_{E}^2\rangle - \langle N_{E}\rangle ^2
\end{equation}
The average occupation of a single electron-dense patch
\begin{eqnarray*}
\langle N_{E}\rangle &=& 0.p_0 + 1.p_1 + 2.p_2 + 3.p_3 + ........\\
                      &=& 0.p_0 + 1.p_0 \exp[- \frac{\Delta E}{k_B T}] + 2.p_0 \exp[- \frac{\Delta E + \Delta E_Q}{k_B T}]
                                            + 3.p_0 \exp[- \frac{\Delta E + 2 \Delta E_Q}{k_B T}]+............\\
                      &=& p_0 \exp[- \frac{\Delta E}{k_B T}]\{1.\exp[- 0.\frac{\Delta E_Q}{k_B T}]+ 2.\exp[- 1.\frac{\Delta E_Q}{k_B
                                            T}]+ 3.\exp[- 2.\frac{ \Delta E_Q}{k_B T}] + ......\}\\
                      &=&p_0 \exp[- \frac{\Delta E}{k_B T}] \times \frac{1}{(1-\exp[- \frac{\Delta E_Q}{k_B T}])^2} \hspace{10mm} [  \displaystyle\sum_{n=1}^{\infty} n x^{n-1}]=\frac{1}{(1-x)^2}]\\
                      &=&\frac{\epsilon \exp[-\frac{ \Delta E}{k_B T}]}{\epsilon + \exp[-\frac{\Delta E}{k_B T}]} \times \frac{1}{\epsilon^2}\\
                      &=&\frac{\exp[-\frac{\Delta E}{k_B T}]}{\epsilon (\epsilon+\exp[-\frac{\Delta E}{k_B T}])}\\
\end{eqnarray*}
and
\begin{eqnarray*}
\langle N^2_{E}\rangle &=& 0^2.p_0 + 1^2.p_1 + 2^2.p_2 + 3^2.p_3 + ........\\
                      &=& 1^2.p_0 \exp[- \frac{\Delta E}{k_B T}] + 2^2.p_0 \exp[- \frac{\Delta E + \Delta E_Q}{k_B T}]
                                            + 3^2.p_0 \exp[- \frac{\Delta E + 2 \Delta E_Q}{k_B T}]+............\\
                      &=& p_0 \exp[- \frac{\Delta E}{k_B T}]\{1^2.\exp[- 0.\frac{\Delta E_Q}{k_B T}]+ 2^2.\exp[- 1.\frac{\Delta E_Q}{k_B
                                            T}]+ 3^2.\exp[- 2.\frac{ \Delta E_Q}{k_B T}] + ......\}\\
                      &=&p_0 \exp[- \frac{\Delta E}{k_B T}] \times \frac{2-\epsilon}{\epsilon^3} \hspace{10mm} [  \displaystyle\sum_{n=1}^{\infty} n^2 x^{n-1}]=\frac{1+x}{(1-x)^3}]\\
                      &=&\frac{(2-\epsilon) \exp[- \frac{\Delta E}{k_B T}]}{\epsilon^2 (\epsilon+\exp[- \frac{\Delta E}{k_B T}])}\\
\end{eqnarray*}
Therefore,
\begin{eqnarray*}
\langle \delta N_{E}^2 \rangle &=& \langle N_{E}^2 \rangle - \langle N_{E} \rangle^2\\
                                &=& \frac{(2-\epsilon) \exp[- \frac{\Delta E}{k_B T}]}{\epsilon^2 (\epsilon+\exp[- \frac{\Delta E}{k_B T}])}- \frac{\exp[-\frac{2\Delta E}{k_B T}]}{\epsilon^2 (\epsilon+\exp[-\frac{\Delta E}{k_B T}])^2}\\
                                &=& \frac{\exp[- \frac{\Delta E}{k_B T}]}{\epsilon^2} \times [ \frac{2-\epsilon}{\epsilon + \exp[- \frac{\Delta E}{k_B T}]}-\frac{\exp[- \frac{\Delta E}{k_B T}]}{(\epsilon + \exp[- \frac{\Delta E}{k_B T}])^2}] \\
                                &=& \frac{\exp[- \frac{\Delta E}{k_B T}]}{\epsilon^2} \times [\frac{(2-\epsilon)(\epsilon + \exp[- \frac{\Delta E}{k_B T}])-\exp[- \frac{\Delta E}{k_B T}]}{(\epsilon + \exp[\frac{-\Delta E}{k_B T}])^2}]\\
                                &=& \frac{\exp[- \frac{\Delta E}{k_B T}]}{\epsilon^2}\times \frac{2 \epsilon + \exp[-\frac{\Delta E}{k_B T}] - \epsilon \exp[\frac{-\Delta E}{k_B T}]}{(\epsilon + \exp[\frac{-\Delta E}{k_B
                                T}])^2}\hspace{5mm} [\mbox{Assuming $\epsilon^2 \rightarrow 0$ in the numerator}]\\
                                &\approx& \frac{\exp[- \frac{\Delta E}{k_B T}]}{\epsilon^2}\times \frac{2 \epsilon}{\epsilon^2}\\
                                &=& \frac{2 \exp[- \frac{\Delta E}{k_B T}]}{\epsilon^3}\\
\end{eqnarray*}

where we assume $ \epsilon > \exp[- \frac{\Delta E}{k_B T}]$, and this approximation is valid till the size of the patches are below 20~nm. We
calculate
\begin{eqnarray*}
\epsilon &=& 1-\exp[-\frac{\Delta E_Q}{k_B T}]\\
         &\approx& 1- [1- \frac{\Delta E_Q}{k_B T}]\hspace{5mm}[\mbox{$\Delta E_Q \ll k_B T$ for patch size more than 5~nm}]\\
         &=& \frac{\Delta E_Q}{k_B T}\\
\end{eqnarray*}
Using Eq.~S8, S9, and $\langle \delta N_{E}^2 \rangle$, we find
\begin{eqnarray*}
\frac{\langle\delta \sigma^2\rangle}{\sigma^2} &=& \frac{\langle\delta N^2\rangle}{N^2}\\
                                               &=& \frac{N_{E}}{N^2} \times \langle \delta N_{E}^2 \rangle \\
                                               &=& 2 \left[\frac{k_B T}{\Delta E_Q}\right] ^3 \frac{N_{E}}{N^2} \exp[- \frac{\Delta E}{k_B T}]\\
\end{eqnarray*}

As we have already shown that
\begin{equation}
\frac{\langle\delta \sigma^2\rangle}{\sigma^2} = |\frac{S_{\sigma}}{\sigma^2}|_{1~Hz}~ ln(f_2/f_1)
\end{equation}
Therefore, we obtain
\begin{equation}
|\frac{S_{\sigma}}{\sigma^2}|_{1~Hz}=\frac{2}{ln(f_2/f_1)} \left[\frac{k_B T}{\Delta E_Q}\right] ^3 \frac{N_{E}}{N^2} \exp[- \frac{\Delta E}{k_B
T}]
\end{equation}

\subsection{Calculation of Hooge parameter~($\gamma_H$) from equation~1 in main text:}
\begin{equation}
\frac{\langle\delta \sigma^2\rangle}{\sigma^2}=2 \left[\frac{k_B T}{\Delta E_Q}\right] ^3 \frac{N_{E}}{N^2} \exp[- \frac{\Delta E}{k_B T}]\\
\end{equation}
\begin{eqnarray*}
\frac{\langle\delta \sigma^2\rangle}{\sigma^2}& =& \int_{f_1}^{f_2} S_{\sigma}/\sigma^2 df \hspace{5mm} [\mbox{$f_1$ and $f_2$ represent
measured frequency bandwidth}]\\
                                              &=& \frac{\gamma_H}{N}\int_{f_1}^{f_2} df/f \hspace{5mm}[\mbox{where we use the Hooge relation
                                              $S_{\sigma}/\sigma^2~=~\gamma_H \frac{1}{Nf}$}]\\
                                              &=& \frac{\gamma_H}{N} ln(f_2/f_1)
\end{eqnarray*}
Therefore,
\begin{equation}
\gamma_H = \frac{2}{ln(f_2/f_1)} \left[\frac{k_B T}{\Delta E_Q}\right] ^3 \frac{N_{E}}{N} \exp[- \frac{\Delta E}{k_B T}]
\end{equation}
This equation predicts $\gamma_H \propto 1/n$. As $\sigma \propto n^2$ at high temperature and gate voltages, the field effect mobility
$\mu_{FE} \propto n$. Therefore, equation~S9 suggests that the Hooge parameter is inversely proportional to $\mu_{FE}$.

If $A$ and $t$ are the area and thickness respectively of the MoS$_2$ channel then
\begin{equation}
N_E = N_{p}\times At =\frac{0.05}{D^2t}At = \frac{0.05A}{D^2}
\end{equation}
We calculate $\Delta E_Q~\simeq~\frac{h^2}{8m^*D^2}~\simeq$~1.6~meV assuming $m^*$ to be roughly equal to free electron mass. Using
$f_1$~=~0.03, $f_2$~=~6.8~Hz and $k_BT$~=~26~meV, we obtain $\gamma_H \approx 0.2$ at carrier density $n~=~4.5\times10^{12}$cm$^{-2}$.


\begin{thebibliography}{55}
\expandafter\ifx\csname natexlab\endcsname\relax\def\natexlab#1{#1}\fi \expandafter\ifx\csname bibnamefont\endcsname\relax
  \def\bibnamefont#1{#1}\fi
\expandafter\ifx\csname bibfnamefont\endcsname\relax
  \def\bibfnamefont#1{#1}\fi
\expandafter\ifx\csname citenamefont\endcsname\relax
  \def\citenamefont#1{#1}\fi
\expandafter\ifx\csname url\endcsname\relax
  \def\url#1{\texttt{#1}}\fi
\expandafter\ifx\csname urlprefix\endcsname\relax\def\urlprefix{URL }\fi \providecommand{\bibinfo}[2]{#2}
\providecommand{\eprint}[2][]{\url{#2}}

\bibitem[{\citenamefont{Radisavljevic
  et~al.}(2011{\natexlab{a}})\citenamefont{Radisavljevic, Radenovic, Brivio,
  Giacometti, and Kis}}]{single}
\bibinfo{author}{\bibfnamefont{B.}~\bibnamefont{Radisavljevic}},
  \bibinfo{author}{\bibfnamefont{A.}~\bibnamefont{Radenovic}},
  \bibinfo{author}{\bibfnamefont{J.}~\bibnamefont{Brivio}},
  \bibinfo{author}{\bibfnamefont{V.}~\bibnamefont{Giacometti}},
  \bibnamefont{and} \bibinfo{author}{\bibfnamefont{A.}~\bibnamefont{Kis}},
  \bibinfo{journal}{Nat. Nano.} \textbf{\bibinfo{volume}{6}},
  \bibinfo{pages}{147} (\bibinfo{year}{2011}{\natexlab{a}}).

\bibitem[{\citenamefont{Liu et~al.}(2012)\citenamefont{Liu, Neal, and
  Ye}}]{channellengthscaling}
\bibinfo{author}{\bibfnamefont{H.}~\bibnamefont{Liu}},
  \bibinfo{author}{\bibfnamefont{A.~T.} \bibnamefont{Neal}}, \bibnamefont{and}
  \bibinfo{author}{\bibfnamefont{P.~D.} \bibnamefont{Ye}},
  \bibinfo{journal}{ACS Nano} \textbf{\bibinfo{volume}{6}},
  \bibinfo{pages}{8563} (\bibinfo{year}{2012}).

\bibitem[{\citenamefont{Radisavljevic
  et~al.}(2011{\natexlab{b}})\citenamefont{Radisavljevic, Whitwick, and
  Kis}}]{kisintegratedcircuit}
\bibinfo{author}{\bibfnamefont{B.}~\bibnamefont{Radisavljevic}},
  \bibinfo{author}{\bibfnamefont{M.~B.} \bibnamefont{Whitwick}},
  \bibnamefont{and} \bibinfo{author}{\bibfnamefont{A.}~\bibnamefont{Kis}},
  \bibinfo{journal}{ACS Nano} \textbf{\bibinfo{volume}{5}},
  \bibinfo{pages}{9934} (\bibinfo{year}{2011}{\natexlab{b}}).

\bibitem[{\citenamefont{Wang et~al.}(2012)\citenamefont{Wang, Yu, Lee, Shi,
  Hsu, Chin, Li, Dubey, Kong, and Palacios}}]{integratedbilayer}
\bibinfo{author}{\bibfnamefont{H.}~\bibnamefont{Wang}},
  \bibinfo{author}{\bibfnamefont{L.}~\bibnamefont{Yu}},
  \bibinfo{author}{\bibfnamefont{Y.-H.} \bibnamefont{Lee}},
  \bibinfo{author}{\bibfnamefont{Y.}~\bibnamefont{Shi}},
  \bibinfo{author}{\bibfnamefont{A.}~\bibnamefont{Hsu}},
  \bibinfo{author}{\bibfnamefont{M.~L.} \bibnamefont{Chin}},
  \bibinfo{author}{\bibfnamefont{L.-J.} \bibnamefont{Li}},
  \bibinfo{author}{\bibfnamefont{M.}~\bibnamefont{Dubey}},
  \bibinfo{author}{\bibfnamefont{J.}~\bibnamefont{Kong}}, \bibnamefont{and}
  \bibinfo{author}{\bibfnamefont{T.}~\bibnamefont{Palacios}},
  \bibinfo{journal}{Nano Letters} \textbf{\bibinfo{volume}{12}},
  \bibinfo{pages}{4674} (\bibinfo{year}{2012}).

\bibitem[{\citenamefont{Yin et~al.}(2012)\citenamefont{Yin, Li, Li, Jiang, Shi,
  Sun, Lu, Zhang, Chen, and Zhang}}]{Yin}
\bibinfo{author}{\bibfnamefont{Z.}~\bibnamefont{Yin}},
  \bibinfo{author}{\bibfnamefont{H.}~\bibnamefont{Li}},
  \bibinfo{author}{\bibfnamefont{H.}~\bibnamefont{Li}},
  \bibinfo{author}{\bibfnamefont{L.}~\bibnamefont{Jiang}},
  \bibinfo{author}{\bibfnamefont{Y.}~\bibnamefont{Shi}},
  \bibinfo{author}{\bibfnamefont{Y.}~\bibnamefont{Sun}},
  \bibinfo{author}{\bibfnamefont{G.}~\bibnamefont{Lu}},
  \bibinfo{author}{\bibfnamefont{Q.}~\bibnamefont{Zhang}},
  \bibinfo{author}{\bibfnamefont{X.}~\bibnamefont{Chen}}, \bibnamefont{and}
  \bibinfo{author}{\bibfnamefont{H.}~\bibnamefont{Zhang}},
  \bibinfo{journal}{ACS Nano} \textbf{\bibinfo{volume}{6}}, \bibinfo{pages}{74}
  (\bibinfo{year}{2012}).

\bibitem[{\citenamefont{Roy et~al.}(2013{\natexlab{a}})\citenamefont{Roy,
  Padmanabhan, Goswami, Phanindra~Sai, Ramalingam, Raghavan, and
  Ghosh}}]{KallolNatNano}
\bibinfo{author}{\bibfnamefont{K.}~\bibnamefont{Roy}},
  \bibinfo{author}{\bibfnamefont{M.}~\bibnamefont{Padmanabhan}},
  \bibinfo{author}{\bibfnamefont{S.}~\bibnamefont{Goswami}},
  \bibinfo{author}{\bibfnamefont{T.}~\bibnamefont{Phanindra~Sai}},
  \bibinfo{author}{\bibfnamefont{G.}~\bibnamefont{Ramalingam}},
  \bibinfo{author}{\bibfnamefont{S.}~\bibnamefont{Raghavan}}, \bibnamefont{and}
  \bibinfo{author}{\bibfnamefont{A.}~\bibnamefont{Ghosh}},
  \bibinfo{journal}{Nat Nano} \textbf{\bibinfo{volume}{8}},
  \bibinfo{pages}{826} (\bibinfo{year}{2013}{\natexlab{a}}).

\bibitem[{\citenamefont{Roy et~al.}(2013{\natexlab{b}})\citenamefont{Roy,
  Padmanabhan, Goswami, Sai, Kaushal, and Ghosh}}]{kallolSSC}
\bibinfo{author}{\bibfnamefont{K.}~\bibnamefont{Roy}},
  \bibinfo{author}{\bibfnamefont{M.}~\bibnamefont{Padmanabhan}},
  \bibinfo{author}{\bibfnamefont{S.}~\bibnamefont{Goswami}},
  \bibinfo{author}{\bibfnamefont{T.~P.} \bibnamefont{Sai}},
  \bibinfo{author}{\bibfnamefont{S.}~\bibnamefont{Kaushal}}, \bibnamefont{and}
  \bibinfo{author}{\bibfnamefont{A.}~\bibnamefont{Ghosh}},
  \bibinfo{journal}{Solid State Communications}
  \textbf{\bibinfo{volume}{175-176}}, \bibinfo{pages}{35}
  (\bibinfo{year}{2013}{\natexlab{b}}).

\bibitem[{\citenamefont{Yoon et~al.}(2011)\citenamefont{Yoon, Ganapathi, and
  Salahuddin}}]{howgood}
\bibinfo{author}{\bibfnamefont{Y.}~\bibnamefont{Yoon}},
  \bibinfo{author}{\bibfnamefont{K.}~\bibnamefont{Ganapathi}},
  \bibnamefont{and}
  \bibinfo{author}{\bibfnamefont{S.}~\bibnamefont{Salahuddin}},
  \bibinfo{journal}{Nano Letters} \textbf{\bibinfo{volume}{11}},
  \bibinfo{pages}{3768} (\bibinfo{year}{2011}).

\bibitem[{\citenamefont{Kaasbjerg et~al.}(2012)\citenamefont{Kaasbjerg,
  Thygesen, and Jacobsen}}]{phononlimitedmobility}
\bibinfo{author}{\bibfnamefont{K.}~\bibnamefont{Kaasbjerg}},
  \bibinfo{author}{\bibfnamefont{K.~S.} \bibnamefont{Thygesen}},
  \bibnamefont{and} \bibinfo{author}{\bibfnamefont{K.~W.}
  \bibnamefont{Jacobsen}}, \bibinfo{journal}{Phys. Rev. B}
  \textbf{\bibinfo{volume}{85}}, \bibinfo{pages}{115317}
  (\bibinfo{year}{2012}).

\bibitem[{\citenamefont{Das et~al.}(2013)\citenamefont{Das, Chen, Penumatcha,
  and Appenzeller}}]{scandium}
\bibinfo{author}{\bibfnamefont{S.}~\bibnamefont{Das}},
  \bibinfo{author}{\bibfnamefont{H.-Y.} \bibnamefont{Chen}},
  \bibinfo{author}{\bibfnamefont{A.~V.} \bibnamefont{Penumatcha}},
  \bibnamefont{and}
  \bibinfo{author}{\bibfnamefont{J.}~\bibnamefont{Appenzeller}},
  \bibinfo{journal}{Nano Letters} \textbf{\bibinfo{volume}{13}},
  \bibinfo{pages}{100} (\bibinfo{year}{2013}).

\bibitem[{\citenamefont{Liu and Ye}(2012)}]{dualgatemos2ieee}
\bibinfo{author}{\bibfnamefont{H.}~\bibnamefont{Liu}} \bibnamefont{and}
  \bibinfo{author}{\bibfnamefont{P.}~\bibnamefont{Ye}},
  \bibinfo{journal}{Electron Device Letters, IEEE}
  \textbf{\bibinfo{volume}{33}}, \bibinfo{pages}{546 } (\bibinfo{year}{2012}).

\bibitem[{\citenamefont{Late et~al.}(2012)\citenamefont{Late, Liu, Matte,
  Dravid, and Rao}}]{cnrraomos2}
\bibinfo{author}{\bibfnamefont{D.~J.} \bibnamefont{Late}},
  \bibinfo{author}{\bibfnamefont{B.}~\bibnamefont{Liu}},
  \bibinfo{author}{\bibfnamefont{H.~S. S.~R.} \bibnamefont{Matte}},
  \bibinfo{author}{\bibfnamefont{V.~P.} \bibnamefont{Dravid}},
  \bibnamefont{and} \bibinfo{author}{\bibfnamefont{C.~N.~R.}
  \bibnamefont{Rao}}, \bibinfo{journal}{ACS Nano} \textbf{\bibinfo{volume}{6}},
  \bibinfo{pages}{5635} (\bibinfo{year}{2012}).

\bibitem[{\citenamefont{Katsnelson and Geim}(2008)}]{corrugation}
\bibinfo{author}{\bibfnamefont{M.}~\bibnamefont{Katsnelson}} \bibnamefont{and}
  \bibinfo{author}{\bibfnamefont{A.}~\bibnamefont{Geim}},
  \bibinfo{journal}{Phil. Trans. R. Soc. A} \textbf{\bibinfo{volume}{366}},
  \bibinfo{pages}{195} (\bibinfo{year}{2008}).

\bibitem[{\citenamefont{Adam et~al.}(2009)\citenamefont{Adam, Hwang, Rossi, and
  Sarma}}]{Adamchargedimpuritysolidstatecomm}
\bibinfo{author}{\bibfnamefont{S.}~\bibnamefont{Adam}},
  \bibinfo{author}{\bibfnamefont{E.}~\bibnamefont{Hwang}},
  \bibinfo{author}{\bibfnamefont{E.}~\bibnamefont{Rossi}}, \bibnamefont{and}
  \bibinfo{author}{\bibfnamefont{S.~D.} \bibnamefont{Sarma}},
  \bibinfo{journal}{Solid State Communications} \textbf{\bibinfo{volume}{149}},
  \bibinfo{pages}{1072 } (\bibinfo{year}{2009}).

\bibitem[{\citenamefont{Chen et~al.}(2008)\citenamefont{Chen, Jang, Adam,
  Fuhrer, Williams, and Ishigami}}]{chargedimpuritygraphene}
\bibinfo{author}{\bibfnamefont{J.-H.} \bibnamefont{Chen}},
  \bibinfo{author}{\bibfnamefont{C.}~\bibnamefont{Jang}},
  \bibinfo{author}{\bibfnamefont{S.}~\bibnamefont{Adam}},
  \bibinfo{author}{\bibfnamefont{M.~S.} \bibnamefont{Fuhrer}},
  \bibinfo{author}{\bibfnamefont{E.~D.} \bibnamefont{Williams}},
  \bibnamefont{and} \bibinfo{author}{\bibfnamefont{M.}~\bibnamefont{Ishigami}},
  \bibinfo{journal}{Nat. Phys.} \textbf{\bibinfo{volume}{4}},
  \bibinfo{pages}{377} (\bibinfo{year}{2008}).

\bibitem[{\citenamefont{Pal et~al.}(2011)\citenamefont{Pal, Ghatak, Kochat,
  Sneha, Sampathkumar, Raghavan, and Ghosh}}]{ouracsnano}
\bibinfo{author}{\bibfnamefont{A.~N.} \bibnamefont{Pal}},
  \bibinfo{author}{\bibfnamefont{S.}~\bibnamefont{Ghatak}},
  \bibinfo{author}{\bibfnamefont{V.}~\bibnamefont{Kochat}},
  \bibinfo{author}{\bibfnamefont{E.~S.} \bibnamefont{Sneha}},
  \bibinfo{author}{\bibfnamefont{A.}~\bibnamefont{Sampathkumar}},
  \bibinfo{author}{\bibfnamefont{S.}~\bibnamefont{Raghavan}}, \bibnamefont{and}
  \bibinfo{author}{\bibfnamefont{A.}~\bibnamefont{Ghosh}},
  \bibinfo{journal}{ACS Nano} \textbf{\bibinfo{volume}{5}},
  \bibinfo{pages}{2075} (\bibinfo{year}{2011}).

\bibitem[{\citenamefont{Ayari et~al.}(2007)\citenamefont{Ayari, Cobas,
  Ogundadegbe, and Fuhrer}}]{nanopatch}
\bibinfo{author}{\bibfnamefont{A.}~\bibnamefont{Ayari}},
  \bibinfo{author}{\bibfnamefont{E.}~\bibnamefont{Cobas}},
  \bibinfo{author}{\bibfnamefont{O.}~\bibnamefont{Ogundadegbe}},
  \bibnamefont{and} \bibinfo{author}{\bibfnamefont{M.~S.}
  \bibnamefont{Fuhrer}}, \bibinfo{journal}{Journal of Applied Physics}
  \textbf{\bibinfo{volume}{101}}, \bibinfo{pages}{014507}
  (\bibinfo{year}{2007}).

\bibitem[{\citenamefont{Ghatak et~al.}(2011)\citenamefont{Ghatak, Pal, and
  Ghosh}}]{natureofelectronic}
\bibinfo{author}{\bibfnamefont{S.}~\bibnamefont{Ghatak}},
  \bibinfo{author}{\bibfnamefont{A.~N.} \bibnamefont{Pal}}, \bibnamefont{and}
  \bibinfo{author}{\bibfnamefont{A.}~\bibnamefont{Ghosh}},
  \bibinfo{journal}{ACS Nano} \textbf{\bibinfo{volume}{5}},
  \bibinfo{pages}{7707} (\bibinfo{year}{2011}).

\bibitem[{\citenamefont{Qiu et~al.}(2013)\citenamefont{Qiu, Xu, Wang, Ren, Nan,
  Ni, Chen, Yuan, Miao, Song et~al.}}]{quisulphurvac}
\bibinfo{author}{\bibfnamefont{H.}~\bibnamefont{Qiu}},
  \bibinfo{author}{\bibfnamefont{T.}~\bibnamefont{Xu}},
  \bibinfo{author}{\bibfnamefont{Z.}~\bibnamefont{Wang}},
  \bibinfo{author}{\bibfnamefont{W.}~\bibnamefont{Ren}},
  \bibinfo{author}{\bibfnamefont{H.}~\bibnamefont{Nan}},
  \bibinfo{author}{\bibfnamefont{Z.}~\bibnamefont{Ni}},
  \bibinfo{author}{\bibfnamefont{Q.}~\bibnamefont{Chen}},
  \bibinfo{author}{\bibfnamefont{S.}~\bibnamefont{Yuan}},
  \bibinfo{author}{\bibfnamefont{F.}~\bibnamefont{Miao}},
  \bibinfo{author}{\bibfnamefont{F.}~\bibnamefont{Song}}, \bibnamefont{et~al.},
  \bibinfo{journal}{Nat Commun} \textbf{\bibinfo{volume}{4}},
  \bibinfo{pages}{2642} (\bibinfo{year}{2013}).

\bibitem[{\citenamefont{Windom et~al.}(2011)\citenamefont{Windom, Sawyer, and
  Hahn}}]{MoO3raman}
\bibinfo{author}{\bibfnamefont{B.}~\bibnamefont{Windom}},
  \bibinfo{author}{\bibfnamefont{W.}~\bibnamefont{Sawyer}}, \bibnamefont{and}
  \bibinfo{author}{\bibfnamefont{D.}~\bibnamefont{Hahn}},
  \bibinfo{journal}{Tribology Letters} \textbf{\bibinfo{volume}{42}},
  \bibinfo{pages}{301} (\bibinfo{year}{2011}).

\bibitem[{\citenamefont{Eda et~al.}(2011)\citenamefont{Eda, Yamaguchi, Voiry,
  Fujita, Chen, and Chhowalla}}]{photolumichemicallyexfo}
\bibinfo{author}{\bibfnamefont{G.}~\bibnamefont{Eda}},
  \bibinfo{author}{\bibfnamefont{H.}~\bibnamefont{Yamaguchi}},
  \bibinfo{author}{\bibfnamefont{D.}~\bibnamefont{Voiry}},
  \bibinfo{author}{\bibfnamefont{T.}~\bibnamefont{Fujita}},
  \bibinfo{author}{\bibfnamefont{M.}~\bibnamefont{Chen}}, \bibnamefont{and}
  \bibinfo{author}{\bibfnamefont{M.}~\bibnamefont{Chhowalla}},
  \bibinfo{journal}{Nano Letters} \textbf{\bibinfo{volume}{11}},
  \bibinfo{pages}{5111} (\bibinfo{year}{2011}).

\bibitem[{\citenamefont{Chandni et~al.}(2009)\citenamefont{Chandni, Ghosh,
  Vijaya, and Mohan}}]{Chandnidiprl}
\bibinfo{author}{\bibfnamefont{U.}~\bibnamefont{Chandni}},
  \bibinfo{author}{\bibfnamefont{A.}~\bibnamefont{Ghosh}},
  \bibinfo{author}{\bibfnamefont{H.~S.} \bibnamefont{Vijaya}},
  \bibnamefont{and} \bibinfo{author}{\bibfnamefont{S.}~\bibnamefont{Mohan}},
  \bibinfo{journal}{Phys. Rev. Lett.} \textbf{\bibinfo{volume}{102}},
  \bibinfo{pages}{025701} (\bibinfo{year}{2009}).

\bibitem[{\citenamefont{Pal and Ghosh}(2009{\natexlab{a}})}]{atindaapl}
\bibinfo{author}{\bibfnamefont{A.~N.} \bibnamefont{Pal}} \bibnamefont{and}
  \bibinfo{author}{\bibfnamefont{A.}~\bibnamefont{Ghosh}},
  \bibinfo{journal}{Applied Physics Letters} \textbf{\bibinfo{volume}{95}},
  \bibinfo{pages}{082105} (\bibinfo{year}{2009}{\natexlab{a}}).

\bibitem[{\citenamefont{Pal and Ghosh}(2009{\natexlab{b}})}]{atindaprl}
\bibinfo{author}{\bibfnamefont{A.~N.} \bibnamefont{Pal}} \bibnamefont{and}
  \bibinfo{author}{\bibfnamefont{A.}~\bibnamefont{Ghosh}},
  \bibinfo{journal}{Phys. Rev. Lett.} \textbf{\bibinfo{volume}{102}},
  \bibinfo{pages}{126805} (\bibinfo{year}{2009}{\natexlab{b}}).

\bibitem[{\citenamefont{Jayaraman and Sodini}(1989)}]{jayaraman}
\bibinfo{author}{\bibfnamefont{R.}~\bibnamefont{Jayaraman}} \bibnamefont{and}
  \bibinfo{author}{\bibfnamefont{C.~G.} \bibnamefont{Sodini}},
  \bibinfo{journal}{IEEE Trans. Electron Device} \textbf{\bibinfo{volume}{36}},
  \bibinfo{pages}{1773} (\bibinfo{year}{1989}).

\bibitem[{\citenamefont{loannidis et~al.}(2000)\citenamefont{loannidis,
  Tsormpatzoglou, Tassis, Dimitriadis, Templier, and Kamarinos}}]{loannidis}
\bibinfo{author}{\bibfnamefont{E.~G.} \bibnamefont{loannidis}},
  \bibinfo{author}{\bibfnamefont{A.}~\bibnamefont{Tsormpatzoglou}},
  \bibinfo{author}{\bibfnamefont{D.~H.} \bibnamefont{Tassis}},
  \bibinfo{author}{\bibfnamefont{C.~A.} \bibnamefont{Dimitriadis}},
  \bibinfo{author}{\bibfnamefont{F.}~\bibnamefont{Templier}}, \bibnamefont{and}
  \bibinfo{author}{\bibfnamefont{G.}~\bibnamefont{Kamarinos}},
  \bibinfo{journal}{J. Appl. Phys.} \textbf{\bibinfo{volume}{108}},
  \bibinfo{pages}{106103} (\bibinfo{year}{2000}).

\bibitem[{\citenamefont{Adam and Das~Sarma}(2008)}]{sigmansqdassarma}
\bibinfo{author}{\bibfnamefont{S.}~\bibnamefont{Adam}} \bibnamefont{and}
  \bibinfo{author}{\bibfnamefont{S.}~\bibnamefont{Das~Sarma}},
  \bibinfo{journal}{Phys. Rev. B} \textbf{\bibinfo{volume}{77}},
  \bibinfo{pages}{115436} (\bibinfo{year}{2008}).

\bibitem[{\citenamefont{Berleb et~al.}(2000)\citenamefont{Berleb, Muckl,
  Brutting, and Schwoerer}}]{Berleb}
\bibinfo{author}{\bibfnamefont{S.}~\bibnamefont{Berleb}},
  \bibinfo{author}{\bibfnamefont{A.~G.} \bibnamefont{Muckl}},
  \bibinfo{author}{\bibfnamefont{W.}~\bibnamefont{Brutting}}, \bibnamefont{and}
  \bibinfo{author}{\bibfnamefont{M.}~\bibnamefont{Schwoerer}},
  \bibinfo{journal}{Synthetic Metals} \textbf{\bibinfo{volume}{111-112}},
  \bibinfo{pages}{341 } (\bibinfo{year}{2000}).

\bibitem[{\citenamefont{de~Boer et~al.}(2004)\citenamefont{de~Boer, Gershenson,
  Morpurgo, and Podzorov}}]{mottgurneymorpugo}
\bibinfo{author}{\bibfnamefont{R.~W.~I.} \bibnamefont{de~Boer}},
  \bibinfo{author}{\bibfnamefont{M.~E.} \bibnamefont{Gershenson}},
  \bibinfo{author}{\bibfnamefont{A.~F.} \bibnamefont{Morpurgo}},
  \bibnamefont{and} \bibinfo{author}{\bibfnamefont{V.}~\bibnamefont{Podzorov}},
  \bibinfo{journal}{physica status solidi (a)} \textbf{\bibinfo{volume}{201}},
  \bibinfo{pages}{1302} (\bibinfo{year}{2004}).

\bibitem[{\citenamefont{Joung et~al.}(2010)\citenamefont{Joung, Chunder, Zhai,
  and Khondaker}}]{khandekerrgo}
\bibinfo{author}{\bibfnamefont{D.}~\bibnamefont{Joung}},
  \bibinfo{author}{\bibfnamefont{A.}~\bibnamefont{Chunder}},
  \bibinfo{author}{\bibfnamefont{L.}~\bibnamefont{Zhai}}, \bibnamefont{and}
  \bibinfo{author}{\bibfnamefont{S.~I.} \bibnamefont{Khondaker}},
  \bibinfo{journal}{Applied Physics Letters} \textbf{\bibinfo{volume}{97}},
  \bibinfo{pages}{093105} (\bibinfo{year}{2010}).

\bibitem[{\citenamefont{Ghatak and Ghosh}(2013)}]{mottgurneyourAPL}
\bibinfo{author}{\bibfnamefont{S.}~\bibnamefont{Ghatak}} \bibnamefont{and}
  \bibinfo{author}{\bibfnamefont{A.}~\bibnamefont{Ghosh}},
  \bibinfo{journal}{Applied Physics Letters} \textbf{\bibinfo{volume}{103}},
  \bibinfo{pages}{122103} (\bibinfo{year}{2013}).

\bibitem[{\citenamefont{Kumar et~al.}(2003)\citenamefont{Kumar, Jain, Kapoor,
  Poortmans, and Mertens}}]{mottgurneykumer}
\bibinfo{author}{\bibfnamefont{V.}~\bibnamefont{Kumar}},
  \bibinfo{author}{\bibfnamefont{S.~C.} \bibnamefont{Jain}},
  \bibinfo{author}{\bibfnamefont{A.~K.} \bibnamefont{Kapoor}},
  \bibinfo{author}{\bibfnamefont{J.}~\bibnamefont{Poortmans}},
  \bibnamefont{and} \bibinfo{author}{\bibfnamefont{R.}~\bibnamefont{Mertens}},
  \bibinfo{journal}{Journal of Applied Physics} \textbf{\bibinfo{volume}{94}},
  \bibinfo{pages}{1283} (\bibinfo{year}{2003}).

\bibitem[{\citenamefont{Shklovskii}(2003)}]{shklovskii}
\bibinfo{author}{\bibfnamefont{B.~I.} \bibnamefont{Shklovskii}},
  \bibinfo{journal}{Phys. Rev. B} \textbf{\bibinfo{volume}{67}},
  \bibinfo{pages}{045201} (\bibinfo{year}{2003}).

\bibitem[{\citenamefont{Radisavljevic and Kis}(2013)}]{metalinsulatormos2}
\bibinfo{author}{\bibfnamefont{B.}~\bibnamefont{Radisavljevic}}
  \bibnamefont{and} \bibinfo{author}{\bibfnamefont{A.}~\bibnamefont{Kis}},
  \bibinfo{journal}{Nat Mater} \textbf{\bibinfo{volume}{12}},
  \bibinfo{pages}{815} (\bibinfo{year}{2013}).

\bibitem[{\citenamefont{Baugher et~al.}(2013)\citenamefont{Baugher, Churchill,
  Yang, and Jarillo-Herrero}}]{pablomos2}
\bibinfo{author}{\bibfnamefont{B.}~\bibnamefont{Baugher}},
  \bibinfo{author}{\bibfnamefont{H.~O.~H.} \bibnamefont{Churchill}},
  \bibinfo{author}{\bibfnamefont{Y.}~\bibnamefont{Yang}}, \bibnamefont{and}
  \bibinfo{author}{\bibfnamefont{P.}~\bibnamefont{Jarillo-Herrero}},
  \bibinfo{journal}{Nano Letters} \textbf{\bibinfo{volume}{13}},
  \bibinfo{pages}{4212} (\bibinfo{year}{2013}).

\bibitem[{\citenamefont{Dutta and Horn}(1981)}]{duttahorn}
\bibinfo{author}{\bibfnamefont{P.}~\bibnamefont{Dutta}} \bibnamefont{and}
  \bibinfo{author}{\bibfnamefont{P.~M.} \bibnamefont{Horn}},
  \bibinfo{journal}{Rev. Mod. Phys.} \textbf{\bibinfo{volume}{53}},
  \bibinfo{pages}{497} (\bibinfo{year}{1981}).

\bibitem[{\citenamefont{McWhorter}(Philadelphia, University of Pennsylvania
  Press, 1957)}]{mcwhorter}
\bibinfo{author}{\bibfnamefont{A.~L.} \bibnamefont{McWhorter}},
  \emph{\bibinfo{title}{Semiconductor Surface Physics}}
  (\bibinfo{year}{Philadelphia, University of Pennsylvania Press, 1957}).

\bibitem[{\citenamefont{Zhang et~al.}(2011)\citenamefont{Zhang, Mendez, and
  Du}}]{xudu}
\bibinfo{author}{\bibfnamefont{Y.}~\bibnamefont{Zhang}},
  \bibinfo{author}{\bibfnamefont{E.~E.} \bibnamefont{Mendez}},
  \bibnamefont{and} \bibinfo{author}{\bibfnamefont{X.}~\bibnamefont{Du}},
  \bibinfo{journal}{ACS Nano} \textbf{\bibinfo{volume}{5}},
  \bibinfo{pages}{8124} (\bibinfo{year}{2011}).

\bibitem[{\citenamefont{Renteria et~al.}(2013)\citenamefont{Renteria, Samnakay,
  Rumyantsev, Goli, Shur, and Balandin}}]{balandinmos2noise}
\bibinfo{author}{\bibfnamefont{J.}~\bibnamefont{Renteria}},
  \bibinfo{author}{\bibfnamefont{R.}~\bibnamefont{Samnakay}},
  \bibinfo{author}{\bibfnamefont{S.~L.} \bibnamefont{Rumyantsev}},
  \bibinfo{author}{\bibfnamefont{P.}~\bibnamefont{Goli}},
  \bibinfo{author}{\bibfnamefont{M.~S.} \bibnamefont{Shur}}, \bibnamefont{and}
  \bibinfo{author}{\bibfnamefont{A.~A.} \bibnamefont{Balandin}},
  \bibinfo{journal}{arXiv.org e-Print archive}
  \textbf{\bibinfo{volume}{arXiv:1312.6868}} (\bibinfo{year}{2013}).

\bibitem[{\citenamefont{Dean et~al.}(2010)\citenamefont{Dean, Young, Meric,
  Lee, Wang, Sorgenfrei, Watanabe, Taniguchi, Kim, Shepard et~al.}}]{dean1}
\bibinfo{author}{\bibfnamefont{C.~R.} \bibnamefont{Dean}},
  \bibinfo{author}{\bibfnamefont{A.~F.} \bibnamefont{Young}},
  \bibinfo{author}{\bibfnamefont{I.}~\bibnamefont{Meric}},
  \bibinfo{author}{\bibfnamefont{C.}~\bibnamefont{Lee}},
  \bibinfo{author}{\bibfnamefont{L.}~\bibnamefont{Wang}},
  \bibinfo{author}{\bibfnamefont{S.}~\bibnamefont{Sorgenfrei}},
  \bibinfo{author}{\bibfnamefont{K.}~\bibnamefont{Watanabe}},
  \bibinfo{author}{\bibfnamefont{T.}~\bibnamefont{Taniguchi}},
  \bibinfo{author}{\bibfnamefont{P.}~\bibnamefont{Kim}},
  \bibinfo{author}{\bibfnamefont{K.~L.} \bibnamefont{Shepard}},
  \bibnamefont{et~al.}, \bibinfo{journal}{Nat. Nano.}
  \textbf{\bibinfo{volume}{5}}, \bibinfo{pages}{722 } (\bibinfo{year}{2010}).

\bibitem[{\citenamefont{Wang et~al.}(2013)\citenamefont{Wang, Luo, Zhang,
  Laskar, Ma, Wu, Rajan, and Lu}}]{mos2noiseOhioGr}
\bibinfo{author}{\bibfnamefont{Y.}~\bibnamefont{Wang}},
  \bibinfo{author}{\bibfnamefont{X.}~\bibnamefont{Luo}},
  \bibinfo{author}{\bibfnamefont{N.}~\bibnamefont{Zhang}},
  \bibinfo{author}{\bibfnamefont{M.~R.} \bibnamefont{Laskar}},
  \bibinfo{author}{\bibfnamefont{L.}~\bibnamefont{Ma}},
  \bibinfo{author}{\bibfnamefont{Y.}~\bibnamefont{Wu}},
  \bibinfo{author}{\bibfnamefont{S.}~\bibnamefont{Rajan}}, \bibnamefont{and}
  \bibinfo{author}{\bibfnamefont{W.}~\bibnamefont{Lu}},
  \bibinfo{journal}{arXiv.org e-Print archive}
  \textbf{\bibinfo{volume}{1310.6484}} (\bibinfo{year}{2013}).

\bibitem[{\citenamefont{Sangwan et~al.}(2013)\citenamefont{Sangwan, Arnold,
  Jariwala, Marks, Lauhon, and Hersam}}]{noisemos2hersem}
\bibinfo{author}{\bibfnamefont{V.~K.} \bibnamefont{Sangwan}},
  \bibinfo{author}{\bibfnamefont{H.~N.} \bibnamefont{Arnold}},
  \bibinfo{author}{\bibfnamefont{D.}~\bibnamefont{Jariwala}},
  \bibinfo{author}{\bibfnamefont{T.~J.} \bibnamefont{Marks}},
  \bibinfo{author}{\bibfnamefont{L.~J.} \bibnamefont{Lauhon}},
  \bibnamefont{and} \bibinfo{author}{\bibfnamefont{M.~C.}
  \bibnamefont{Hersam}}, \bibinfo{journal}{Nano Letters}
  \textbf{\bibinfo{volume}{13}}, \bibinfo{pages}{4351} (\bibinfo{year}{2013}).

\bibitem[{\citenamefont{McDonnell et~al.}(0)\citenamefont{McDonnell, Addou,
  Buie, Wallace, and Hinkle}}]{defectmediated}
\bibinfo{author}{\bibfnamefont{S.}~\bibnamefont{McDonnell}},
  \bibinfo{author}{\bibfnamefont{R.}~\bibnamefont{Addou}},
  \bibinfo{author}{\bibfnamefont{C.}~\bibnamefont{Buie}},
  \bibinfo{author}{\bibfnamefont{R.~M.} \bibnamefont{Wallace}},
  \bibnamefont{and} \bibinfo{author}{\bibfnamefont{C.~L.}
  \bibnamefont{Hinkle}}, \bibinfo{journal}{ACS Nano}
  \textbf{\bibinfo{volume}{doi:10.1021/nn500044q}}, \bibinfo{pages}{null}
  (\bibinfo{year}{0}).








  \bibitem[{\citenamefont{Benameur et~al.}(2011)\citenamefont{Benameur,
  Radisavljevic, Héron, Sahoo, Berger, and Kis}}]{visibilitykis}
\bibinfo{author}{\bibfnamefont{M.~M.} \bibnamefont{Benameur}},
  \bibinfo{author}{\bibfnamefont{B.}~\bibnamefont{Radisavljevic}},
  \bibinfo{author}{\bibfnamefont{J.~S.} \bibnamefont{Héron}},
  \bibinfo{author}{\bibfnamefont{S.}~\bibnamefont{Sahoo}},
  \bibinfo{author}{\bibfnamefont{H.}~\bibnamefont{Berger}}, \bibnamefont{and}
  \bibinfo{author}{\bibfnamefont{A.}~\bibnamefont{Kis}},
  \bibinfo{journal}{Nanotechnology} \textbf{\bibinfo{volume}{22}},
  \bibinfo{pages}{125706} (\bibinfo{year}{2011}).

\bibitem[{\citenamefont{Castellanos-Gomez
  et~al.}(2010)\citenamefont{Castellanos-Gomez, Agrait, and
  Rubio-Bollinger}}]{visibilitygomez}
\bibinfo{author}{\bibfnamefont{A.}~\bibnamefont{Castellanos-Gomez}},
  \bibinfo{author}{\bibfnamefont{N.}~\bibnamefont{Agrait}}, \bibnamefont{and}
  \bibinfo{author}{\bibfnamefont{G.}~\bibnamefont{Rubio-Bollinger}},
  \bibinfo{journal}{Applied Physics Letters} \textbf{\bibinfo{volume}{96}},
  \bibinfo{pages}{213116} (\bibinfo{year}{2010}).

\bibitem[{\citenamefont{Lee et~al.}(2010)\citenamefont{Lee, Yan, Brus, Heinz,
  Hone, and Ryu}}]{Ramanacsnano}
\bibinfo{author}{\bibfnamefont{C.}~\bibnamefont{Lee}},
  \bibinfo{author}{\bibfnamefont{H.}~\bibnamefont{Yan}},
  \bibinfo{author}{\bibfnamefont{L.~E.} \bibnamefont{Brus}},
  \bibinfo{author}{\bibfnamefont{T.~F.} \bibnamefont{Heinz}},
  \bibinfo{author}{\bibfnamefont{J.}~\bibnamefont{Hone}}, \bibnamefont{and}
  \bibinfo{author}{\bibfnamefont{S.}~\bibnamefont{Ryu}}, \bibinfo{journal}{ACS
  Nano} \textbf{\bibinfo{volume}{4}}, \bibinfo{pages}{2695}
  (\bibinfo{year}{2010}).

\bibitem[{\citenamefont{Zomer et~al.}(2011)\citenamefont{Zomer, Dash, Tombros,
  and van Wees}}]{weesgroup}
\bibinfo{author}{\bibfnamefont{P.~J.} \bibnamefont{Zomer}},
  \bibinfo{author}{\bibfnamefont{S.~P.} \bibnamefont{Dash}},
  \bibinfo{author}{\bibfnamefont{N.}~\bibnamefont{Tombros}}, \bibnamefont{and}
  \bibinfo{author}{\bibfnamefont{B.~J.} \bibnamefont{van Wees}},
  \bibinfo{journal}{Applied Physics Letters} \textbf{\bibinfo{volume}{99}},
  \bibinfo{pages}{232104} (\bibinfo{year}{2011}).

\bibitem[{\citenamefont{Scofield}(1987)}]{scofield1byfnoise}
\bibinfo{author}{\bibfnamefont{J.~H.} \bibnamefont{Scofield}},
  \bibinfo{journal}{Review of Scientific Instruments}
  \textbf{\bibinfo{volume}{58}}, \bibinfo{pages}{985} (\bibinfo{year}{1987}).

\bibitem[{\citenamefont{Ghosh et~al.}(2008)\citenamefont{Ghosh, Kar, Bid, and
  Raychaudhuri}}]{arindamdetailnoisearxiv}
\bibinfo{author}{\bibfnamefont{A.}~\bibnamefont{Ghosh}},
  \bibinfo{author}{\bibfnamefont{S.}~\bibnamefont{Kar}},
  \bibinfo{author}{\bibfnamefont{A.}~\bibnamefont{Bid}}, \bibnamefont{and}
  \bibinfo{author}{\bibfnamefont{A.~K.} \bibnamefont{Raychaudhuri}},
  \bibinfo{journal}{arXiv.org e-Print archive}
  \textbf{\bibinfo{volume}{arXiv:cond-mat/0402130}} (\bibinfo{year}{2008}).

\bibitem[{\citenamefont{Sarma et~al.}(2000)\citenamefont{Sarma, Mahadevan,
  Saha-Dasgupta, Ray, and Kumar}}]{DDxpsPRL}
\bibinfo{author}{\bibfnamefont{D.~D.} \bibnamefont{Sarma}},
  \bibinfo{author}{\bibfnamefont{P.}~\bibnamefont{Mahadevan}},
  \bibinfo{author}{\bibfnamefont{T.}~\bibnamefont{Saha-Dasgupta}},
  \bibinfo{author}{\bibfnamefont{S.}~\bibnamefont{Ray}}, \bibnamefont{and}
  \bibinfo{author}{\bibfnamefont{A.}~\bibnamefont{Kumar}},
  \bibinfo{journal}{Phys. Rev. Lett.} \textbf{\bibinfo{volume}{85}},
  \bibinfo{pages}{2549} (\bibinfo{year}{2000}).

\bibitem[{\citenamefont{Sarma and Rao}(1980)}]{DDandCNRRao}
\bibinfo{author}{\bibfnamefont{D.~D.} \bibnamefont{Sarma}} \bibnamefont{and}
  \bibinfo{author}{\bibfnamefont{C.~N.~R.} \bibnamefont{Rao}},
  \bibinfo{journal}{Journal of Electron Spectroscopy and Related Phenomena}
  \textbf{\bibinfo{volume}{20}}, \bibinfo{pages}{25 } (\bibinfo{year}{1980}).


\bibitem[{\citenamefont{Ye et~al.}(2012)\citenamefont{Ye, Zhang, Akashi,
  Bahramy, Arita, and Iwasa}}]{superconductingdome}
\bibinfo{author}{\bibfnamefont{J.~T.} \bibnamefont{Ye}},
  \bibinfo{author}{\bibfnamefont{Y.~J.} \bibnamefont{Zhang}},
  \bibinfo{author}{\bibfnamefont{R.}~\bibnamefont{Akashi}},
  \bibinfo{author}{\bibfnamefont{M.~S.} \bibnamefont{Bahramy}},
  \bibinfo{author}{\bibfnamefont{R.}~\bibnamefont{Arita}}, \bibnamefont{and}
  \bibinfo{author}{\bibfnamefont{Y.}~\bibnamefont{Iwasa}},
  \bibinfo{journal}{Science} \textbf{\bibinfo{volume}{338}},
  \bibinfo{pages}{1193} (\bibinfo{year}{2012}).


\bibitem[{\citenamefont{Pradhan et~al.}(2013)\citenamefont{Pradhan, Rhodes,
  Zhang, Talapatra, Terrones, Ajayan, and Balicas}}]{intrinsicajayan}
\bibinfo{author}{\bibfnamefont{N.~R.} \bibnamefont{Pradhan}},
  \bibinfo{author}{\bibfnamefont{D.}~\bibnamefont{Rhodes}},
  \bibinfo{author}{\bibfnamefont{Q.}~\bibnamefont{Zhang}},
  \bibinfo{author}{\bibfnamefont{S.}~\bibnamefont{Talapatra}},
  \bibinfo{author}{\bibfnamefont{M.}~\bibnamefont{Terrones}},
  \bibinfo{author}{\bibfnamefont{P.~M.} \bibnamefont{Ajayan}},
  \bibnamefont{and} \bibinfo{author}{\bibfnamefont{L.}~\bibnamefont{Balicas}},
  \bibinfo{journal}{Applied Physics Letters} \textbf{\bibinfo{volume}{102}},
  \bibinfo{pages}{123105} (\bibinfo{year}{2013}).

\bibitem[{\citenamefont{Fuhrer and Hone}(2013)}]{mobilitydualgatefuhrer}
\bibinfo{author}{\bibfnamefont{M.~S.} \bibnamefont{Fuhrer}} \bibnamefont{and}
  \bibinfo{author}{\bibfnamefont{J.}~\bibnamefont{Hone}}, \bibinfo{journal}{Nat
  Nano} \textbf{\bibinfo{volume}{8}}, \bibinfo{pages}{146}
  (\bibinfo{year}{2013}).

\bibitem[{\citenamefont{Radisavljevic and
  Kis}(2013{\natexlab{b}})}]{mobilitydualgatekis}
\bibinfo{author}{\bibfnamefont{B.}~\bibnamefont{Radisavljevic}}
  \bibnamefont{and} \bibinfo{author}{\bibfnamefont{A.}~\bibnamefont{Kis}},
  \bibinfo{journal}{Nat Nano} \textbf{\bibinfo{volume}{8}},
  \bibinfo{pages}{147} (\bibinfo{year}{2013}{\natexlab{b}}).

\end{thebibliography}
\end{document}